\documentclass[11pt]{iopart}

\usepackage[british]{babel}
\usepackage{color}
\usepackage{mathrsfs}
\usepackage{setstack}
\usepackage{iopams}
\usepackage[T1]{fontenc}
\usepackage{graphicx}
\usepackage{cite}

\begin{document}

\title{Unexpected crossovers in correlated random-diffusivity processes}
\author{Wei Wang$^{\dagger,\ddagger}$, Flavio Seno$^{\sharp}$, Igor M.
Sokolov$^{\flat}$, Aleksei V. Chechkin$^{\ddagger,\S}$, and Ralf
Metzler$^{\ddagger}$}

\address{$\dagger$ College of Aerospace Engineering, Nanjing University of
Aeronautics and Astronautics, 210016 Nanjing, China\\
$\ddagger$ Institute for Physics \& Astronomy, University of Potsdam, 14476
Potsdam-Golm, Germany\\
$\sharp$ INFN, Padova Section, and Department of Physics and Astronomy "G.
Galilei", University of Padova, Via Marzolo 8 35131, Padova, Italy\\
$\flat$ Institute of Physics, Humboldt University Berlin, Newtonstrasse 15,
D-12489 Berlin, Germany\\
$\S$ Akhiezer Institute for Theoretical Physics, Kharkov 61108, Ukraine}
\ead{rmetzler@uni-potsdam.de (Corresponding author: R Metzler)}

\begin{abstract}
The passive and active motion of micron-sized tracer particles in crowded liquids
and inside living biological cells is ubiquitously characterised by "viscoelastic"
anomalous diffusion, in which the increments of the motion feature long-ranged
negative and positive correlations. While viscoelastic anomalous diffusion is
typically modelled by a Gaussian process with correlated increments, so-called
fractional Gaussian noise, an increasing number of systems are reported, in which
viscoelastic anomalous diffusion is paired with non-Gaussian displacement
distributions. Following recent advances in Brownian yet non-Gaussian diffusion we
here introduce and discuss several possible versions of random-diffusivity models
with long-ranged correlations. While all these models show a crossover from
non-Gaussian to Gaussian distributions beyond some correlation time, their mean
squared displacements exhibit strikingly different behaviours: depending on the
model crossovers from anomalous to normal diffusion are observed, as well as
unexpected dependencies of the effective diffusion coefficient on the correlation
exponent. Our observations of the strong non-universality of random-diffusivity
viscoelastic anomalous diffusion are important for the analysis of experiments
and a better understanding of the physical origins of "viscoelastic yet
non-Gaussian" diffusion.
\end{abstract}

\section{Introduction}

Gaussianity is so fundamentally engrained in statistics that we almost take it
for granted. The law of large numbers, merging into the central limit theorem
(CLT) states that the sum of independent and identically distributed random
variables with finite variance necessarily converges to a Gaussian ("normal")
limit distribution. A prime example is the Gaussian probability density function
(PDF) $P(x,t)=(4\pi Dt)^{-1/2}\exp(-x^2/[4Dt])$ of Brownian motion that also
encodes the mean squared displacement (MSD) $\langle x^2(t)\rangle=2Dt$ with
the diffusion coefficient $D$ \cite{vankampen}.

The powerful CLT notwithstanding, a growing number of "Brownian yet non-Gaussian"
processes are being reported. The original case was made by the Granick group
for colloid motion along nanotubes and tracer diffusion in gels \cite{granick}.
Similar behaviour is found for nanoparticle diffusion in nanopost arrays \cite{post},
diffusion of colloidal particles on fluid interfaces \cite{xue:BYNG8}, and colloid
motion on membranes as well as in suspension \cite{Goldstein:BYNG9}. For further
examples see \cite{prx,andrey}. Typically, the shape of the PDF in these cases is
exponential ("Laplace distribution"), while in some cases a crossover from
exponential to Gaussian is observed beyond some correlation time
\cite{granick}.\footnote{It is quite likely that in the other examples a similar
crossover to a Gaussian PDF is simply beyond the measurement window.} An invariant
exponential PDF can be explained by "superstatistics" in which the measured PDF
is viewed as an ensemble average over the Gaussian PDSs of individual particles, 
weighted by a diffusivity distribution $p(D)$ \cite{granick,beck}. The crossover
to a Gaussian can be described by the "diffusing-diffusivity" (DD) picture, in which
the diffusion coefficient is assumed to be stochastically varying in time. The
inherent correlation time of the stationary diffusivity process then determines
the crossover of the PDF to a Gaussian at long times whose width is determined by
an "effective" diffusion coefficient. Different versions of DD models have been
discussed, all encoding a short time non-Gaussian and long time Gaussian PDF
\cite{gary,jain,prx,tyagi,vittoria,denis,diego,denis1}. Brownian yet non-Gaussian
dynamics was also derived from extreme value arguments \cite{eli} and for a model
with ongoing tracer multimerisation \cite{eli1}. Several random-diffusivity models
based on Brownian motion were discussed in \cite{zhenya,vittoria1}.

Micron-sized tracers in crowded \emph{in vitro\/} liquids \cite{weiss,lene1}, inside
live biological cells \cite{lene,weiss1,weber,christine}, and lipids in bilayer
membranes \cite{kneller} perform "viscoelastic" anomalous diffusion with MSD
$\langle x^2(t)\rangle\simeq D_Ht^{2H}$ and Hurst exponent $0<H<1/2$. A
hallmark of viscoelastic diffusion is the anticorrelation of the passive tracer
motion \cite{goychuk,weiss,lene1,lene,weiss1,weber,kneller,christine}.\footnote{We
use the term "viscoelastic" to distinguish the long-range correlated anomalous
diffusion considered here from other anomalous diffusion processes such as
continuous time random walks or scaled Brownian motion \cite{pccp}.}
Viscoelastic diffusion at equilibrium is described by the fractional Langevin
equation \cite{lutz,goychuk,pccp}, while in the non-equilibrium of live cells
the description is typically based on fractional Brownian motion (FBM)
\cite{mandelbrot,qian,pccp}. Active, superdiffusive particle transport in live cells
is captured by positively correlated FBM dynamics and Hurst exponent $1/2<H<2$
\cite{christine,samu_jcp}. FBM by definition is a Gaussian process, that is, the
underlying fractional Gaussian noise has a Gaussian amplitude distribution
\cite{mandelbrot,qian}. Yet in a number of systems characterised by viscoelastic
anomalous diffusion it was shown that the tracer particle PDF is non-Gaussian,
including tracer motion in live bacteria and yeast cells \cite{lampo}, protein
diffusion in lipid bilayer membranes \cite{ilpo,membranes} as well as in active
vesicle transport in amoeba cells \cite{samu_jcp}. For invariant non-Gaussian
shapes of the PDF "viscoelastic yet non-Gaussian" diffusion can be modelled by
generalised superstatistics \cite{lampo,jakub}. Yet in the above sub- and
superdiffusive systems we expect the PDF to cross over to Gaussian statistics
beyond some system-specific correlation time. The description of this phenomenon
is the goal of this paper.

With experimental techniques such as \emph{in vivo\/} single-particle tracking,
experimentalists routinely obtain ever more precise insights into molecular
processes in biological cells, e.g., how single proteins are produced and diffusive
to their target \cite{xie,gratton}, how cargo such as messenger RNA molecules
or vesicles are transported \cite{jae_nc,robert,caspi,christine,samu_jcp}, or how
viruses reach the nucleus of an infected cell \cite{seisenhuber}. Such data
allow us to extend models for gene regulation \cite{otto} or motor-based
transport \cite{tolya} and ultimately allow more accurate predictions for viral
infectious pathways, drug delivery, or gene silencing techniques in live cells
or in other complex liquids.

We here address the immanent question for a minimal model of non-Gaussian
viscoelastic diffusion with finite correlation time. Analysing different
extensions of Brownian DD models, now fuelled by correlated Gaussian noise,
we demonstrate that the similarity of these models in the Brownian case
disappears in the anomalous diffusion case. We present detailed results for
this non-universality in the viscoelastic anomalous diffusion case in terms
of the time evolution of the MSDs, the effective diffusivities, and the PDFs
of these processes. Specifically, we show that in some cases anomalous
diffusion persists beyond the correlation time while in others normal
diffusion emerges. Comparing our theoretical predictions with experiments
 will allow us to pinpoint more precisely the exact mechanisms of viscoelastic
yet non-Gaussian diffusion with its high relevance to crowded liquids and live
cells.

\section{FBM-generalisation of the minimal diffusing-diffusivity model}

We first analyse the FBM-generalisation of our minimal
DD model \cite{prx}, whose Langevin equation for the particle position reads
\begin{equation}
\label{eq-1}
dx/dt=\sqrt{2D(t)}\xi_H(t)
\end{equation}
in dimensionless form (see \ref{sec-1}). The dynamics of $D(t)$
is assumed to follow the square of an auxiliary Ornstein-Uhlenbeck process $Y(t)$
\cite{prx},
\begin{eqnarray} 
\label{eq-1b}
D(t)=Y^2(t),\quad dY/dt=-Y+\eta(t).
\end{eqnarray}
In the above $\xi_H(t)$ represents fractional Gaussian noise, understood as the
derivative of smoothed FBM with zero mean and autocovariance $\langle\xi^2_H
\rangle_\tau\equiv\langle\xi_H(t)\xi_H(t+\tau)\rangle$ \cite{mandelbrot,qian}
\begin{equation}
\label{eq-2}
\left\langle\xi^2_H\right\rangle_\tau=(2\delta^2)^{-1}\left(|\tau+\delta|^{2H}
-2|\tau|^{2H}+|\tau-\delta|^{2H}\right),
\end{equation}
decaying as $\langle\xi^2_H\rangle_\tau\sim H(2H-1)\tau^{2H-2}$ for $\tau$ longer
than the physically infinitesimal (smoothening) time scale $\delta$
\cite{mandelbrot}. $\eta(t)$ is a zero-mean white Gaussian noise of unit variance.
We assume equilibrium initial conditions for $Y(t)$,
i.e., $Y(0)$ is taken randomly from the equilibrium distribution $f_{\mathrm{eq}}(
Y)=\pi^{-1/2}\exp(-Y^2)$ \cite{prx,vittoria}. Thus the process $Y(t)$ is stationary
with variance $\langle Y^2\rangle=\langle D\rangle=1/2$. The autocorrelation is
$\langle Y(t)Y(t+\tau)\rangle=\exp(-|\tau|)/2$ with unit correlation time in our
dimensionless units. From equation (\ref{eq-1}) we obtain the MSD (see \ref{sec-2})
\begin{equation}
\label{eq-3}
\langle x^2(t)\rangle=4\int_0^t(t-\tau)K(\tau)\langle\xi^2_H\rangle_\tau d\tau
\end{equation}
with kernel $K(\tau)=\langle\sqrt{D(t_1)D(t_2)}\rangle=(1/\pi)[b(\tau)+a(\tau)
\mathrm{arctan}(a(\tau)/b(\tau))]$, where $\tau=|t_1-t_2|$, $a(\tau)=e^{-\tau}$,
and $b(\tau)=\sqrt{1-a^2(\tau)}$.

\begin{figure*}
\includegraphics[width=5.2cm]{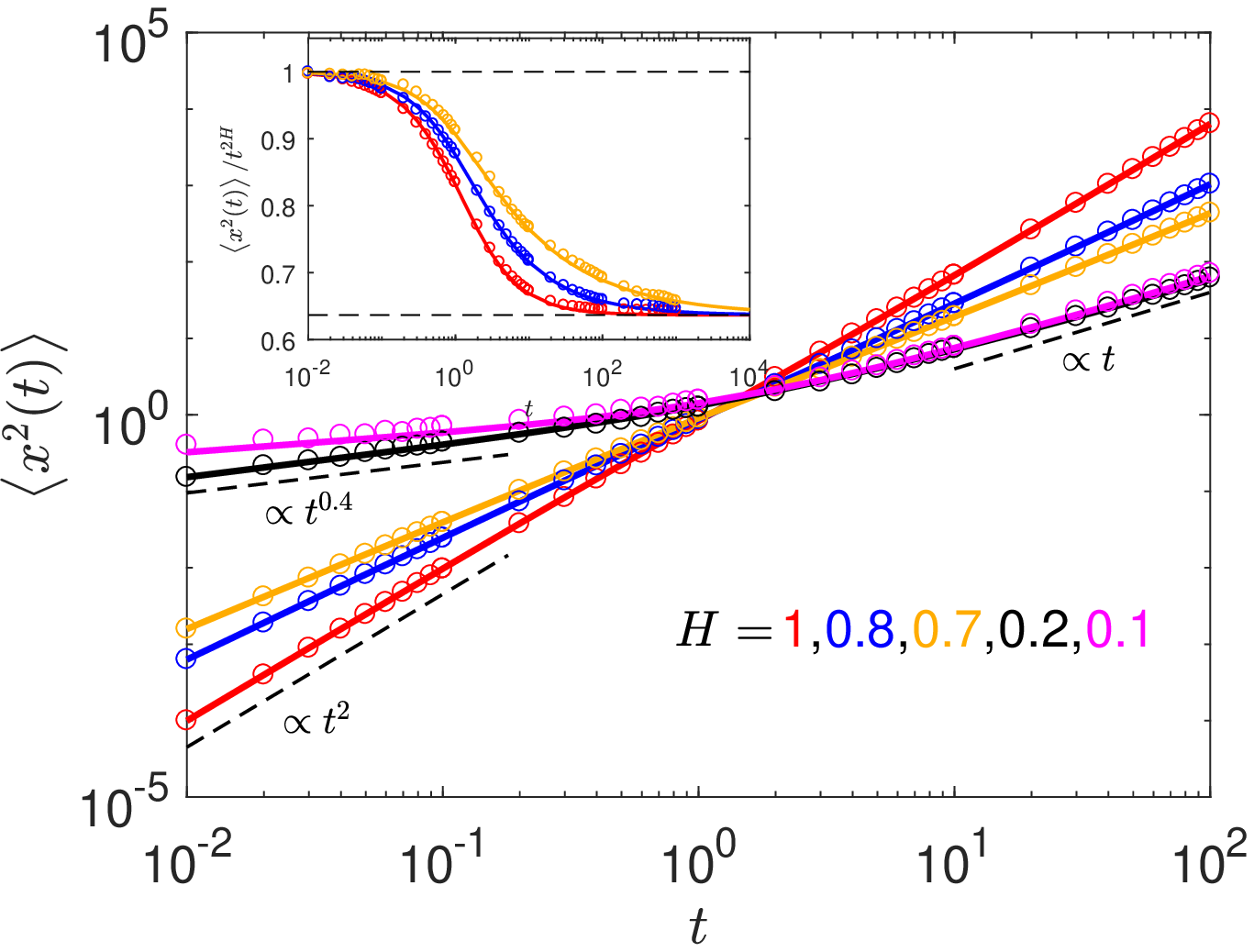}
\includegraphics[width=5.2cm]{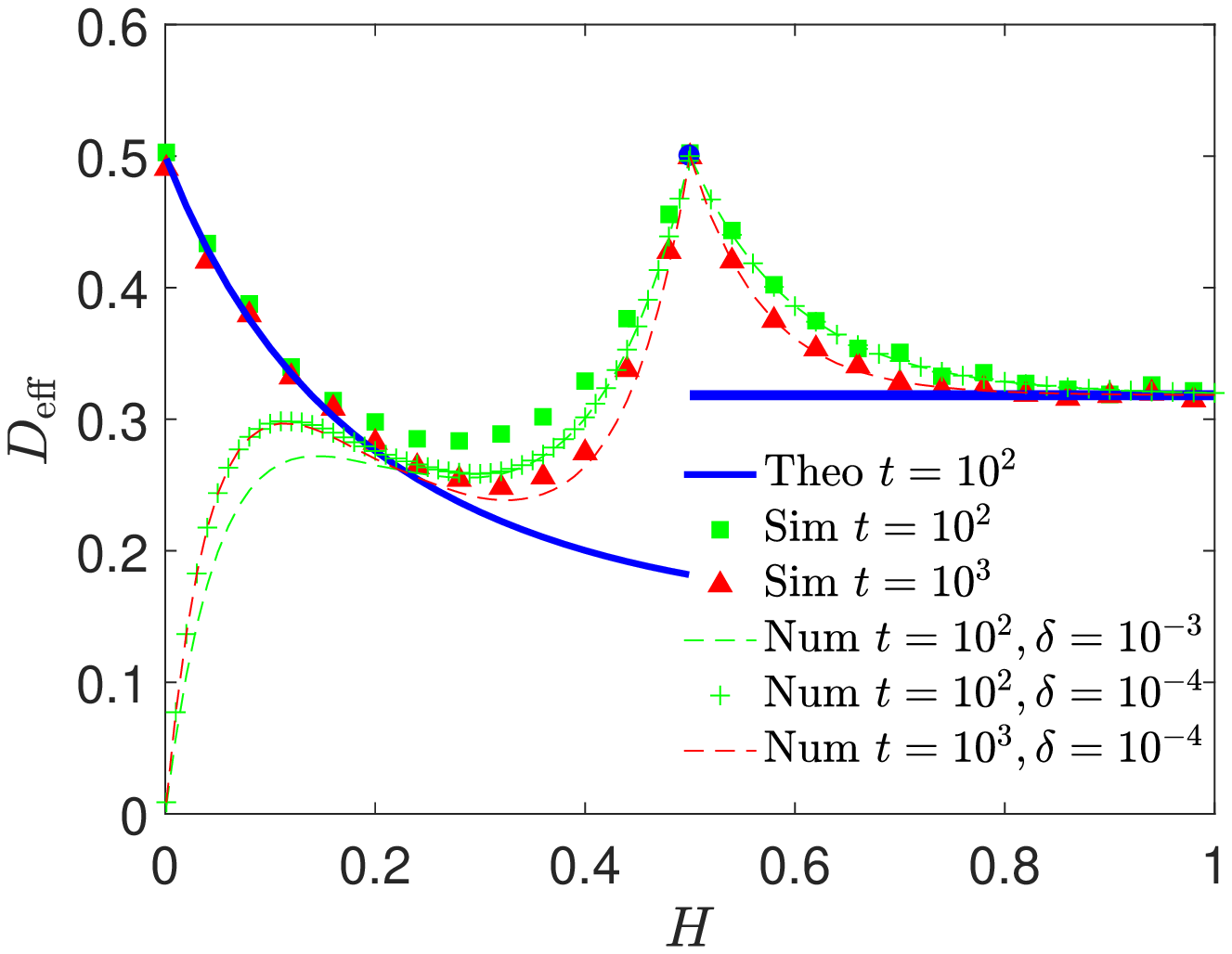}
\includegraphics[width=5.2cm]{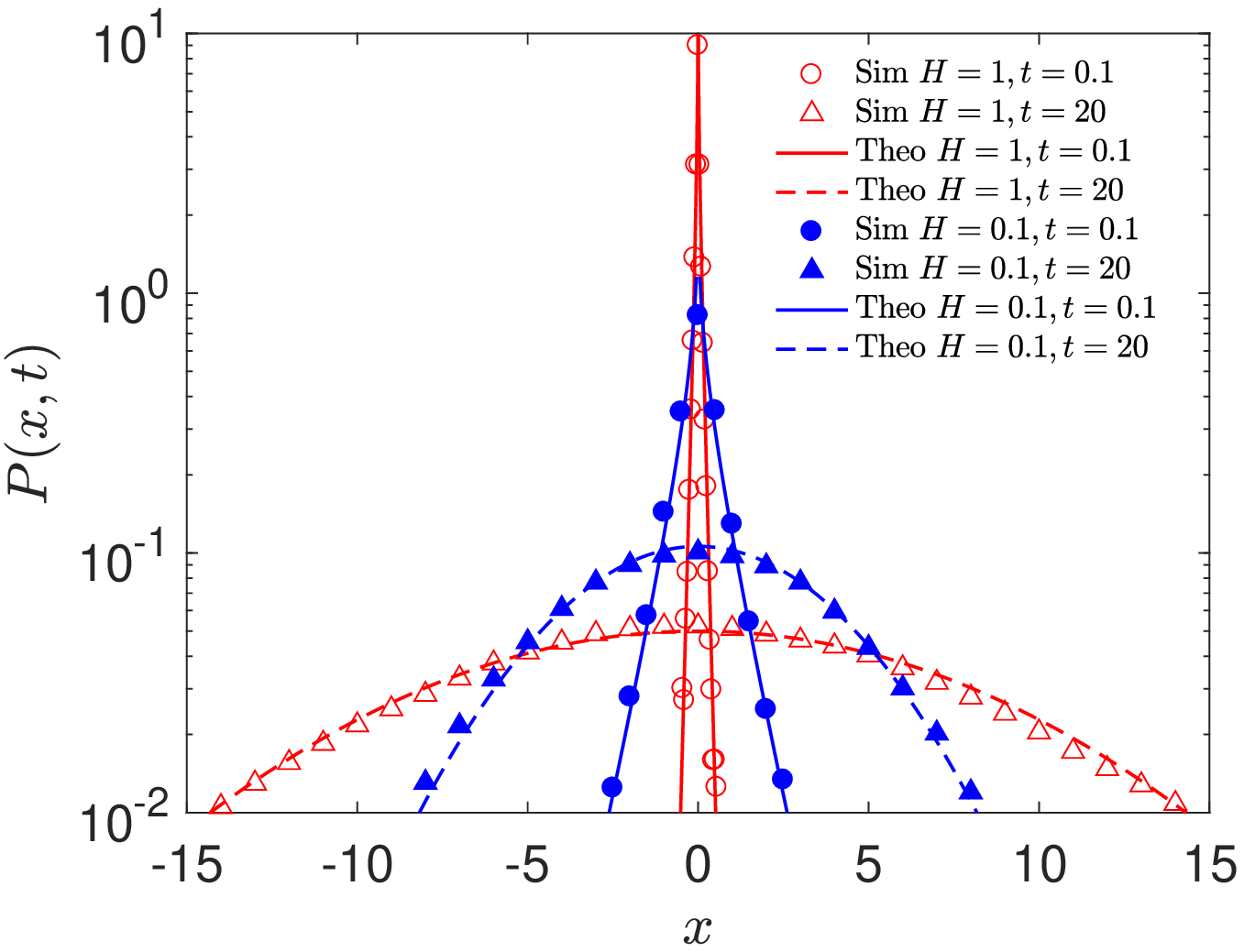}
\caption{Dynamics of the FBM-DD model. Left: Comparison between simulations of
equation (\ref{eq-1}) and the exact MSD (\ref{seq-22}) for $H=1$ as well as
numerical integration of (\ref{eq-3}) for different $H$. The MSD approaches the limits
(dashed lines) $t^{2H}$ at short times and, at long times, anomalous [$(2/\pi)t^{
2H}$] or normal [$2D_{\mathrm{eff}}t$] scaling for super- and subdiffusion,
respectively. Middle: Effective diffusion coefficient as function of $H$. The
theoretical curve (equation (\ref{seq-32}) for $H<1/2$ and $1/\pi$ for $H>1/2$)
shows a distinct discontinuity at the Brownian value $H=1/2$. Results from numerical
evaluation of equations (\ref{seq-23}), (\ref{eq-3}), and simulations are shown to
gradually approach the theoretical values (see text and \ref{sec-4}). Right: Crossover
of the PDF from short-time non-Gaussian form with exponential tails to long-time
Gaussian. The crossover is described in terms of the kurtosis (see text).}
\label{fig1}
\end{figure*}

We first demonstrate how to get the main results for the MSD from simple estimates
at short and long times compared to the correlation time of $D(t)$ dynamics. As
the diffusion coefficient does not change considerably at times shorter than the
correlation time, $K(0)\approx\langle\sqrt{D(t)D(t)}\rangle=\langle D\rangle=
1/2$, equation (\ref{eq-3}) yields
\begin{eqnarray} \label{eq-5}
\langle x^2(t)\rangle\sim4\langle D\rangle\int_0^t(t-\tau)\langle\xi^2_H\rangle
_\tau d\tau=t^{2H}.
\end{eqnarray}
For long times $t\gg1$, more care is needed: as we will see, the long-time limit
is different for the persistent and anti-persistent cases. For the persistent case
$H>1/2$ we assume that the main contribution to the integral in equation (\ref{eq-3}) at
long times comes from large $\tau$, since the noise autocorrelation decays very
slowly. We thus approximate $K(\tau)\approx\langle|Y(t)|\rangle\langle|Y(t+\tau)|
\rangle=\langle|Y(t)|\rangle^2=1/\pi$. Then,
\begin{eqnarray} \label{eq-6}
\langle x^2(t)\rangle\sim4\langle|Y(t)|\rangle^2\int_0^t(t-\tau)\langle\xi^2_H
\rangle_\tau=(2/\pi)t^{2H}.
\end{eqnarray}
In the anti-persistent case $H<1/2$ we split equation (\ref{eq-3}) into two integrals,
$4t\int_0^tK(\tau)\langle\xi^2_H\rangle_\tau d\tau$ and $-4\int_0^t\tau K(\tau)
\langle\xi^2_H\rangle_\tau d\tau$. In the first integral at long times it is
eligible to replace the upper limit of the integral by infinity, since it
converges.\footnote{If the diffusivity is constant, then $K(\tau)$ is constant as
well, and this approximation cannot be used, since necessarily $\int_0^{\infty}
\langle\xi^2_H \rangle_\tau d\tau=0$ in the antipersistent case.}
The second integral produces a subleading term, since it
is bounded from above by $Ct^{2H}$, $C$ being a constant. We therefore have the
following asymptotic result for the MSD in the anti-persistent case at long times,
\begin{eqnarray}
\label{eq-8}
\langle x^2(t)\rangle\sim2D_{\mathrm{eff}}t,
\end{eqnarray}
with $D_{\mathrm{eff}}=\lim_{\delta\to 0}2\int_0^{+\infty}K(\tau)\langle\xi^2_H
\rangle_\tau d\tau$. Thus, the FBM-DD model demonstrates surprising
crossovers in the behaviour of the MSD. In the persistent case the MSD
scales as $t^{2H}$ at both short and long times, but with different diffusion
coefficients. This is in a sharp contrast with the Brownian yet non-Gaussian
diffusion characterised by the \emph{same\/}, invariant diffusivity at all times.
In the antipersistent case the situation is even more counterintuitive: the
subdiffusive scaling of the MSD at short times crosses over to normal diffusion
at long times.

The behaviour of the MSD is shown in figure \ref{fig1}. For superdiffusion, the
change of the diffusivity between the short and long time superdiffusive
scaling $\simeq t^{2H}$ is distinct. Excellent agreement is observed between the
exact and numerical evaluation for $H=1$ and $H=0.7,0.8$, respectively. The exact
analytical expression for $H=1$ is derived in \ref{sec-3}. In the
subdiffusive case simulations and numerical evaluation nicely coincide and show
the crossover from subdiffusion to normal diffusion. Figure \ref{fig1} also shows
the effective long time diffusivity. For superdiffusion the constant value $2/\pi
\approx0.63$ is distinct from the $H$-dependency for subdiffusion ($H<1/2$). For
the Brownian case, $D_{\mathrm{eff}}=1/2$, leading to a distinct
discontinuity at $H=1/2$. Note the slow convergence to the theory of
simulations results and numerical evaluation of the respective integrals
(see \ref{sec-4} for details).

Given the above arguments that at short times ($t\ll1$) the diffusivity is
approximately constant, we expect that in this regime the PDF corresponds to the
superstatistical average of a single Gaussian over the stationary diffusivity
distribution of the Ornstein-Uhlenbeck process,
\begin{eqnarray}
\label{eq-14}
P(x,t)=(\pi\langle x^2(t)\rangle_{\mathrm{ST}}^{1/2})^{-1}K_0(x/\langle x^2(t)\rangle
_{\mathrm{ST}}^{1/2}),
\end{eqnarray}
where $\langle x^2(t)\rangle_{\mathrm{ST}}=t^{2H}$ and $K_0$ is the modified
Bessel function of the second kind \cite{prx}. For long times ($t\gg1$) the
diffusivity correlations decay and the Gaussian limit $P(x,t)=G(\langle x^2(t)
\rangle_{\mathrm{LT}})$ is recovered, where
\begin{eqnarray}
\label{eq-15}
G(\langle x^2(t)\rangle)=(2\pi\langle x^2(t)\rangle)^{-1/2}\exp\left(-x^2/[2
\langle x^2(t)\rangle]\right).
\end{eqnarray}
For $H>1/2$, the long-time MSD is $\langle x^2(t)\rangle_{\mathrm{LT}}=(2/\pi)t^{2H}$
while for $H<1/2$, $\langle x^2(t)\rangle_{\mathrm{LT}}=2D_{\mathrm{eff}}t$. The
crossover behaviour of $P(x,t)$ is indeed corroborated in figure \ref{fig1} for
different values of $H$.

How do these observations compare to generalisations of other established
random-diffusivity models? While in the normal-diffusive regime these models
encode very similar behaviour, we show now that striking differences in the
dynamics emerge when the motion is governed by long-range correlations.

\section{FBM-generalisation of the Tyagi-Cherayil (TC) model}

The generalised TC model \cite{tyagi} in dimensionless units reads
\begin{equation}
\label{ftc}
dx/dt=\sqrt{2}Z(t)\xi_H(t),\,\,\,\,dZ/dt=-Z(t)+\eta(t).
\end{equation}
This expression is obtained from the original equations (\ref{seq-33}) via the
transformations $t\to t/\tau_c$ and $x\to x/(\sigma_1\sigma_2\tau_c^{H+1/2})$.
Using the same notation as before, $\eta$ represents zero-mean white Gaussian
noise and $\xi_H(t)$ is fractional Gaussian noise with Hurst exponent $H$. 

The TC model looks quite similar to the minimal DD model, however, there exists
a decisive difference: In equations (\ref{ftc}) the OU-process $Z(t)$ enters
without the absolute value used in the minimal DD model (\ref{eq-1}). In
expression (\ref{ftc}) the prefactor $Z(t)$ is therefore not a diffusion
coefficient (by definition, a non-negative quantity). In the case $H=1/2$, the
noise $\xi_{1/2}(t)$ is white, that is, uncorrelated, and has zero mean. Due to
the symmetry of the OU-process (for symmetric initial condition) and the noise
$\xi_{1/2}(t)$, the absolute value of $Z(t)$ can be treated as the diffusion
coefficient. In the fractional case, we may still treat $|Z(T)|$ as a diffusion
coefficient, however, in the correlated (persistent or antipersistent) cases we
would then imply a compulsory change in the sign of $\xi(t)$ when $Z(t)$ changes
sign. Yet this model is principally different from the formulation in (\ref{ftc}).
As our discussion shows, the close similarity between the TS and DD models in the
case $H=1/2$ is replaced by a distinct dissimilarity in the emerging dynamics.

The MSD of the FBM-TC model reads
\begin{equation}
\label{seq-34}
\langle x^2(t)\rangle=4\int_0^t(t-\tau)K(\tau)\langle\xi^2_H\rangle_\tau d\tau,
\end{equation}
where the kernel $K(\tau)$ is defined as
\begin{equation}
\label{kernel_tc}
K(\tau)=\langle Z(t_1)Z(t_2)\rangle=\exp(-\tau)/2.
\end{equation}
It is shown in figure \ref{figs1} along with the corresponding Langevin simulations.

Before presenting the exact solution, let us apply an analogous reasoning for the
behaviour of the MSD as developed for the FBM-DD model above. Namely, at short
times we approximate $K(\tau)\approx\langle Z^2\rangle=1/2$. Then equation
(\ref{seq-34}) becomes $\langle x^2(t)\rangle\approx4\langle D\rangle\int_0^t(t
-\tau)\langle\xi_H^2\rangle_{\tau}d\tau\propto t^{2H}$. At long times the MSD can be
composed of the two parts $\langle x^2(t)\rangle=4t\int_0^t K(\tau)\langle\xi_H^2
\rangle_{\tau}d\tau-4\int_0^t\tau K(\tau)\langle\xi_H^2\rangle_{\tau}d\tau$. The
upper limit of the first integral can be replaced by infinity because the first
integral converges in both persistent and anti-persistent cases at long times ($K(
\tau)$ decays to 0 exponentially, different from the FBM-DD model). The second
term is subleading in comparison to the first term. As a result the MSD at long
times scales linearly in time, $\langle x^2(t)\rangle\sim2D_{\mathrm{eff}}t$, for
both sub- and superdiffusion, where $D_{\mathrm{eff}}=\lim_{\delta\to0}2\int_0^{
\infty}K(\tau)\langle\xi_H^2\rangle_{\tau}d\tau$.

\begin{figure*}
\includegraphics[width=5.2cm]{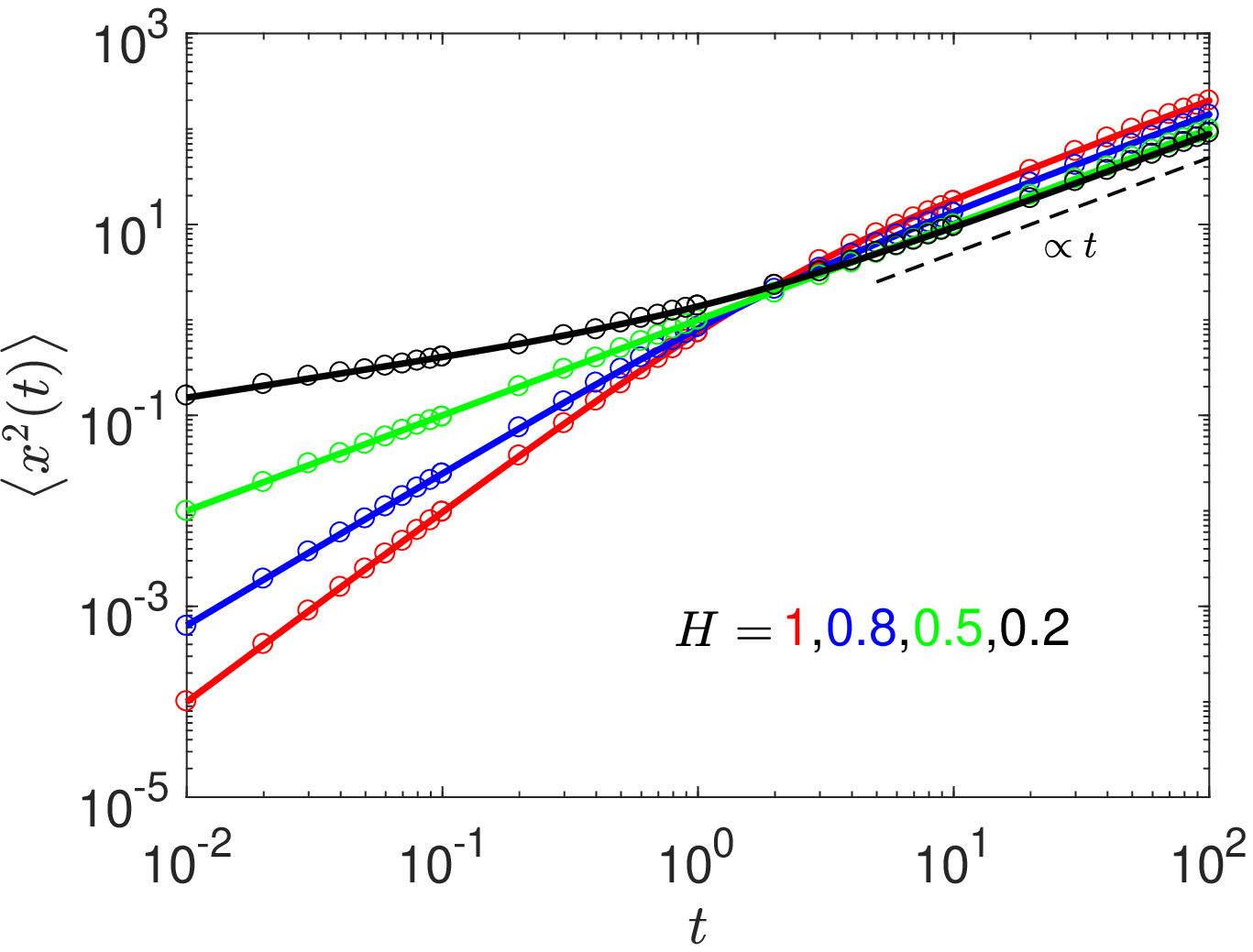}
\includegraphics[width=5.2cm]{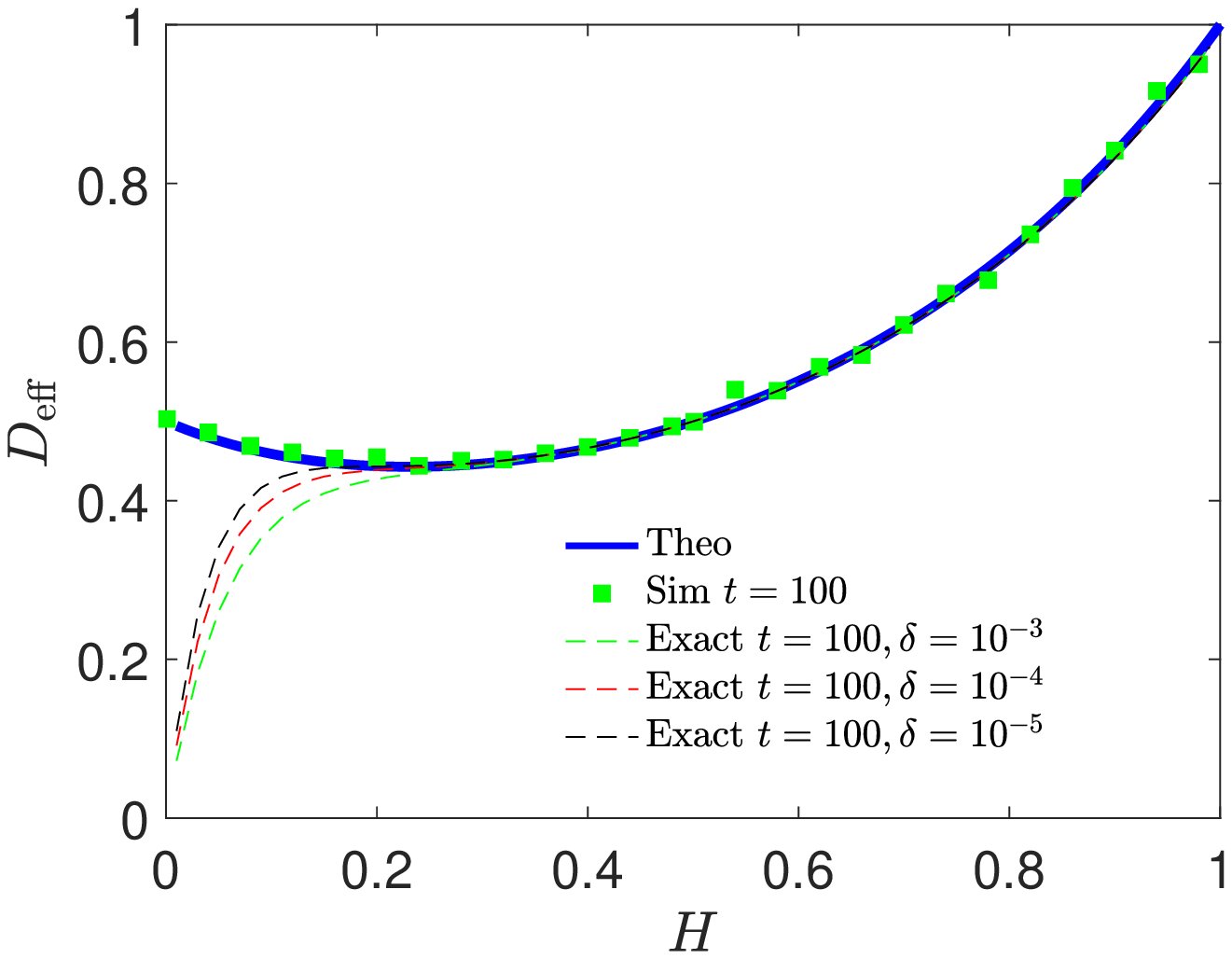}
\includegraphics[width=5.2cm]{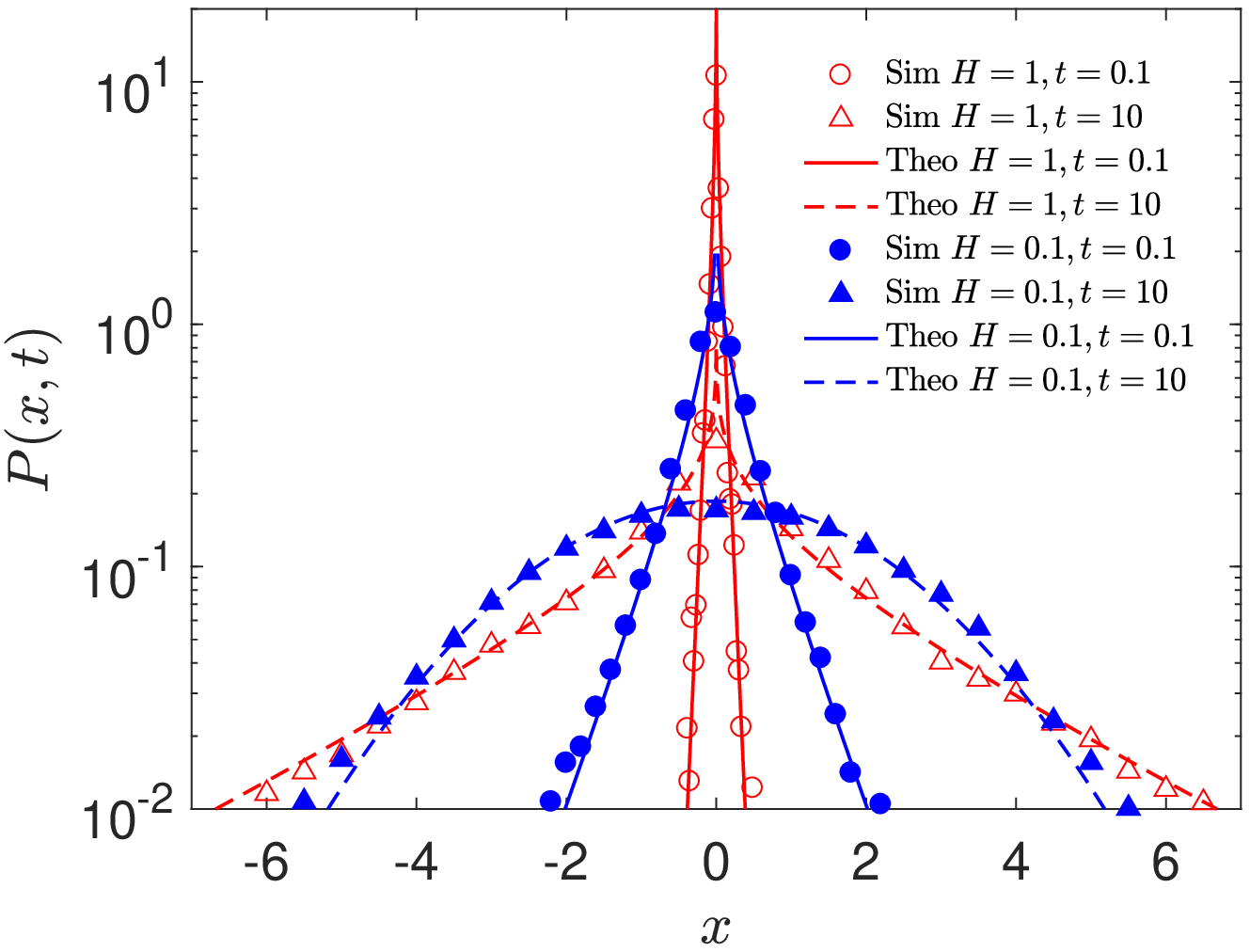}
\caption{Dynamics of the FBM-TC model. Left: Crossover dynamics of the MSD,
showing simulations (symbols) and the solution (\ref{seq-38}). For both sub- and
superdiffusion the long-time scaling is Brownian. Middle: Continuous variation
of the effective diffusion coefficient with $H$. The exact result (\ref{seq-42})
gradually converges
to the theoretical curve for different $\delta$ and $t$. Right: Crossover of the
PDF from short-time non-Gaussian shape with exponential tails to a long-time
Gaussian. The crossover is described in terms of the kurtosis in \ref{sec-5}.}
\label{fig2}
\end{figure*}

Indeed, from the exact form of the MSD in \ref{sec-5} we obtain the limiting
behaviours
\begin{equation}
\label{eq-47}
\langle x^2(t)\rangle\sim\left\{\begin{array}{ll}t^{2H},& t\to0\\
\Gamma(2H+1)t,& t\to\infty\end{array}\right..
\end{equation}
Thus for both sub- and superdiffusion this model shows a crossover from anomalous
to normal diffusion, as demonstrated in figure \ref{fig2}.
The effective long-time diffusion coefficient in this model varies continuously
as $D_{\mathrm{eff}}=\Gamma(2H+1)/2$ for all $H$. In particular, this means that
for $H=1/2$, $D_{\mathrm{eff}}=1/2$. Figure \ref{fig2} shows the exact match of
the simulations results and the numerical evaluation at finite integration step.

The PDF at short times coincides with the superstatistical limit in expression
(\ref{eq-14}) above, as shown explicitly in equation (\ref{seq-43}). At long times
we recover the Gaussian $P(x,t)=G(\Gamma(2H+1)t)$. Note that for $H=1$ the noise is
equal to unity at all times and the dynamics of $x(t)$ is completely determined by
the superstatistic encoded by the OU-process $Z(t)$. The tails of the PDF are thus
always exponential, reflected by the fact that the kurtosis has the invariant value
9 (see \ref{sec-5}).

Despite the strong similarity between the DD and TC models in the Brownian case, for
correlated driving noise their detailed behaviour is strikingly dissimilar, due to
the different asymptotic forms of the kernel $K(\tau)$ (figure \ref{figs1}).

\section{FBM-generalisation of the Switching (S) model}

The third case model we consider here is the S-model with generalised noise
\cite{diego},
\begin{equation}
\label{swit}
dx/dt=\sqrt{2}\theta(t)\xi_H(t),\,\,\,\,\theta(t)=[D_2^{1/2}-D_1^{1/2}]n(t)
+D_1^{1/2},
\end{equation}
where $n(t)$ is a two-state Markov chain switching between the values $\{0,1\}$
and $\xi_H(t)$ represents again fractional Gaussian noise. The constants $D_i$
are the diffusivities in the two states. The switching rates are $k_{12}$ and
$k_{21}$, such that the correlation time is $\tau_c=1/(k_{12}+k_{21})$.

From the first and second moments of the process $\theta(t)$, equations
(\ref{seq-53}) and (\ref{seq-54}), we calculate the MSD of the process. In
the Brownian limit $H=1/2$ the MSD has a linear dependence at all times,
\begin{equation}
\label{seq-55}
\langle x^2(t)\rangle=2(k_{21}D_1+k_{12}D_2)\tau_ct.
\end{equation}
This result was also obtained in \cite{denis1}. For the general case with the
correlation function based on fractional Gaussian noise, we have
\begin{eqnarray}
\nonumber
\langle x^2(t)\rangle&=&4\int_0^t(t-\tau)K(\tau)\langle\xi^2_H\rangle_\tau d\tau\\
\nonumber
&=&2a_1e^{-t/\tau_c}t^{2H}+2a_2t^{2H}+4Ha_1\tau_c^{2H-1}\gamma(2H,t/\tau_c)t\\
\nonumber
&&+2(1-2H)a_1\tau_c^{2H}\gamma(2H+1,t/\tau_c)\\
&&-\frac{2a_1(t+1)\delta^{2H}}{(H+1)(2H+1)}-\frac{2a_2\delta^{2H}}{(H+1)(2H+1)}
+o(\delta^{2H}),
\label{seq-56}
\end{eqnarray}
where $K(\tau)=\left\langle \theta(t_1)\theta(t_2)\right\rangle$, $a_1=(D_2^{1/2}
-D_1^{1/2})^2k_{12}k_{21}\tau_c^2$, and $a_2=(k_{21}D_1^{1/2}+k_{12}D_2^{1/2})^2
\tau_c^2$. At short times $t\ll\tau_c$ we find the scaling behaviour
\begin{equation}
\label{seq-57}
\langle x^2(t)\rangle_{\mathrm{ST}}\sim2(a_1+a_2)t^{2H}=2(k_{21}D_1+k_{12}D_2)
\tau_ct^{2H}.
\end{equation}
At long times $t\gg\tau_c$ the same scaling law is obtained, but with a different
prefactor for the persistent case ($H>1/2$),
\begin{equation}
\label{seq-58}
\left\langle x^2(t)\right\rangle_{\mathrm{LT}}\sim 2a_2t^{2H}=2[k_{21}D_1^{1/2}
+k_{12}D_2^{1/2}]^2\tau_c^2t^{2H}.
\end{equation} 
In contrast, for the anti-persistent case ($H<1/2$), we derive a crossover to
normal diffusion,
\begin{equation}
\label{msd_switch1}
\left\langle x^2(t)\right\rangle_{\mathrm{LT}}\simeq2D_{\mathrm{eff}}t 
\langle x^2(t)\rangle\sim 2[\Gamma(2H+1)\tau^{2H-1}a_1]t.
\end{equation} 
From equations (\ref{seq-55}), (\ref{seq-58}), and (\ref{msd_switch1}), the long-time
effective diffusivity can be obtained as
\begin{eqnarray}
\label{seq-60}
D_{\mathrm{eff}}=\left\{\begin{array}{ll}
(k_{21}D_1^{1/2}+k_{12}D_2^{1/2})^2\tau_c^2,&1/2<H\le1\\[0.2cm]
(k_{21}D_1+k_{12}D_2)\tau_c,&H=1/2\\[0.2cm]
\Gamma(2H+1)(D_2^{1/2}-D_1^{1/2})^2k_{12}k_{21}\tau_c^{2H+1},&0<H< 1/2
\end{array}\right..
\end{eqnarray}

\begin{figure*}
\includegraphics[width=5.2cm]{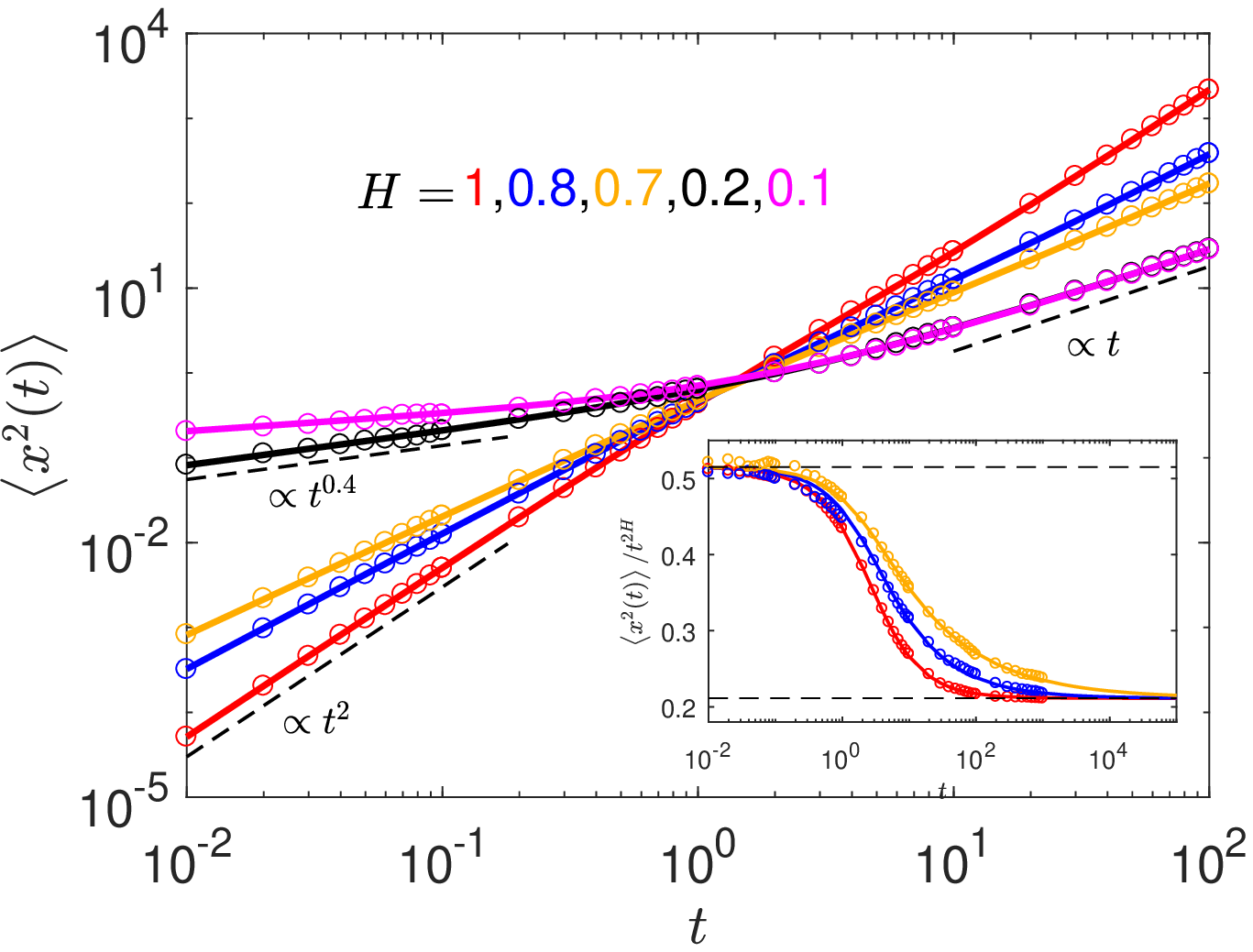}
\includegraphics[width=5.2cm]{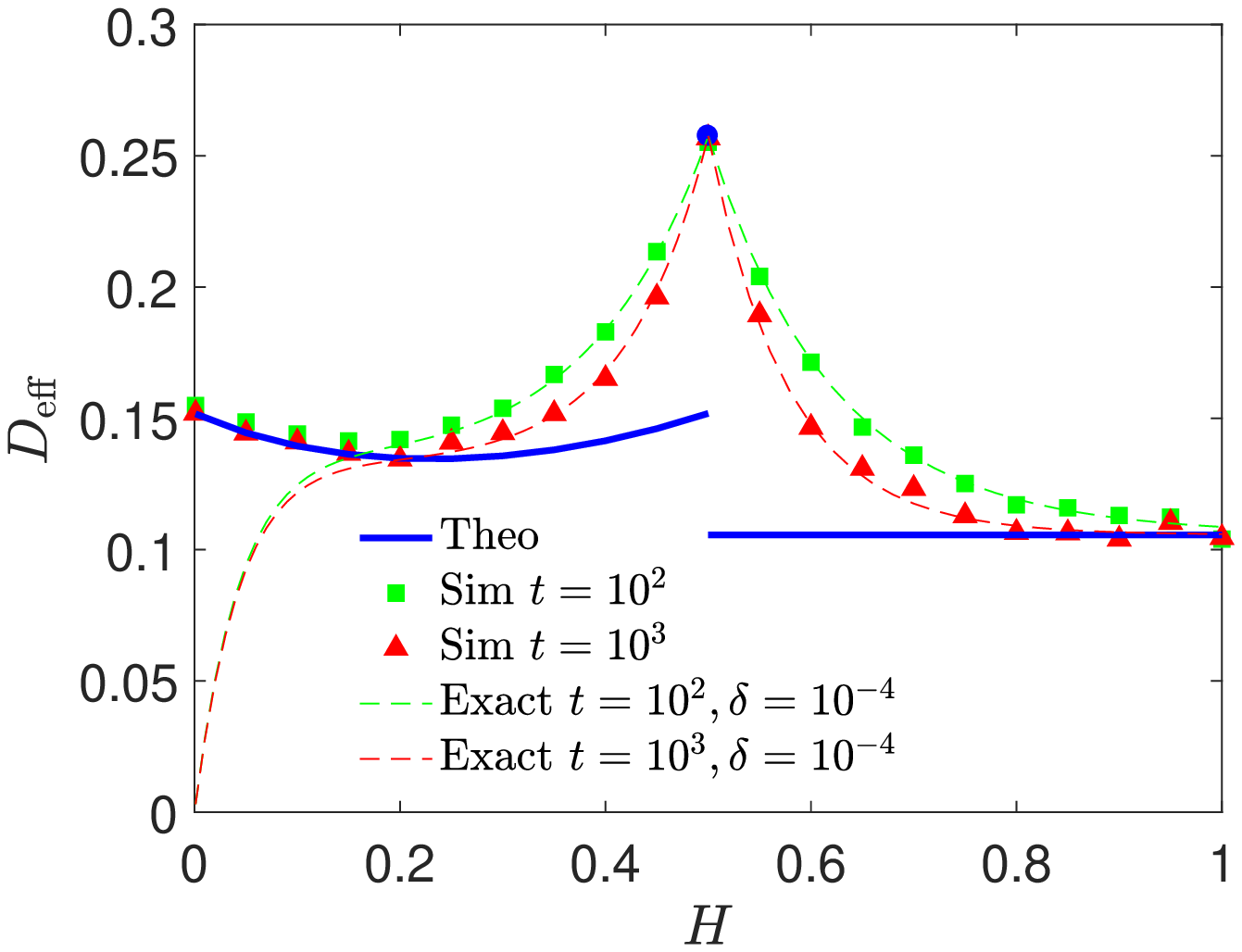}
\includegraphics[width=5.2cm]{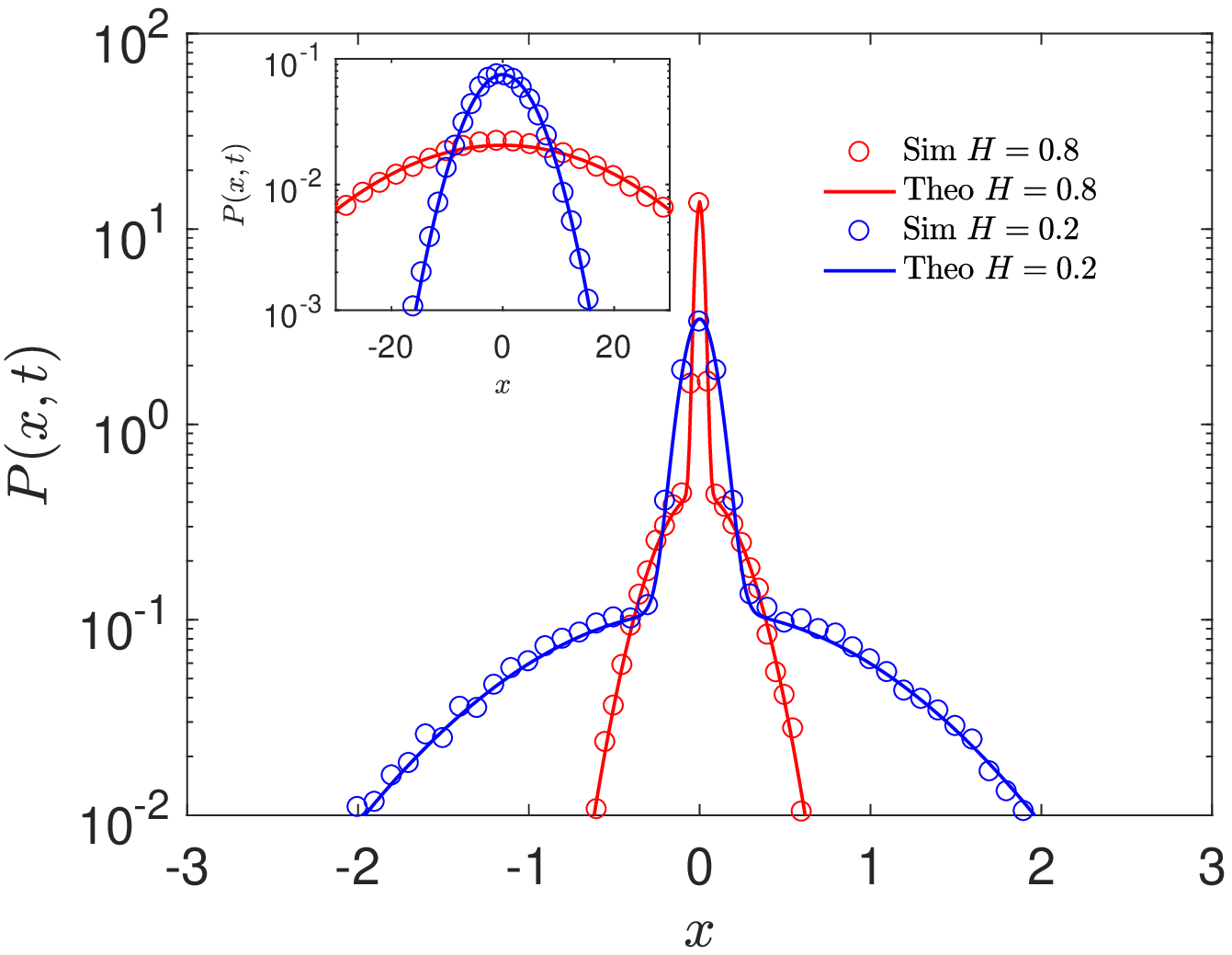}
\caption{Dynamics of the FBM-S model. Left: crossover dynamics of the MSD showing
simulations (symbols) and result (\ref{seq-56}). For sub- and superdiffusion,
respectively, the long-time scaling is $t^{2H}$ and $2D_{\mathrm{eff}}t$. Middle:
$D_{\mathrm{eff}}$ as function of $H$. The theoretical behaviour (\ref{seq-60}) shows
a distinct discontinuity
around the Brownian case $H=1/2$. The gradual convergence of simulations and results
(\ref{seq-61}) and (\ref{seq-62}) for different $t$ and $\delta$ are shown. Right:
Initially the non-Gaussian shape is composed of two Gaussians with different
diffusivity. At long times a single Gaussian emerges. The crossover is described in
terms of the kurtosis (see \ref{sec-6}).}
\label{fig3}
\end{figure*}

The crossover behaviours of the MSD in the peristent and anti-persistent cases,
analogous to the difference in the long-time scalings of the FBM-DD model, are
displayed in figure \ref{fig3}. We also see some similarities between the FBM-S
and FBM-DD models for the effective diffusivity. For the TC-DD model an $H$-dependent
behaviour for $H<1/2$ is followed by a discontinuity at $H=1/2$ and then a constant
value for $H>1/2$. The results of the MSD for finite values $\delta$ and $t$ are
given in \ref{sec-6}.

Next we discuss the PDF and kurtosis. At short times the continuous superstatistic
of the previous cases is reduced to the discrete case of two superimposed Gaussians,
producing the non-exponential form
\begin{eqnarray}
\nonumber
P(x,t)&=&P(x,t|\theta(t)=D_1^{1/2})\times\mathrm{Pr}\{\theta(t)=D_1^{1/2}\}\\
\nonumber
&&+P(x,t|\theta(t)=D_2^{1/2})\times\mathrm{Pr}\{\theta(t)=D_2^{1/2}\}\\
&=&[k_{21}G(2D_1t^{2H})+k_{12}G(2D_2t^{2H})]\tau_c.
\label{seq-63}
\end{eqnarray}
At long times a single Gaussian dominates,
\begin{equation}
\label{seq-64}
P(x,t)\sim G\left(\langle x^2(t)\rangle_{\mathrm{LT}}\right),
\end{equation}
where $\langle x^2(t)\rangle_{\mathrm{LT}}$ is given by equations (\ref{seq-58}) and
(\ref{msd_switch1}) for the super- and subdiffusive cases, respectively.
Figure \ref{fig3} shows the superimposed two Gaussians at short times and the single
Gaussian at long times.

\section{Conclusions}

Viscoelastic anomalous diffusion with
long-ranged correlations is a non-Markovian, natively Gaussian process widely
observed in complex liquids and the cytoplasm of biological cells. Most data
analyses have concentrated on the MSD and the displacement autocorrelation
function. Yet, once probed, the PDF in
many of these systems turns out to be non-Gaussian, a phenomenon ascribed to the
heterogeneity of the systems. Building on recent results for Brownian yet
non-Gaussian diffusion, in which the non-Gaussian ensemble behaviour is
understood as a consequence of a heterogeneous diffusivity coefficient, we here
analysed three different random-diffusivity models driven by
correlated Gaussian noise.

Despite the simplicity of these models we observed surprising
behaviours. Thus, while in the Brownian case all models display a linear PSD with
invariant diffusion coefficient, in the correlated case a crossover occurs from
short to long-time behaviours, with respect to the intrinsic correlation times.
In particular, whether the long-time scaling of the MSD is anomalous or normal,
depends on the specific model. Moreover, the effective diffusivity exhibits
unexpectedly complex behaviours with discontinuities in the FBM-DD and FBM-S
models.

In all cases a crossover from an initial non-Gaussian to a Gaussian PDF occurs.
We showed that the FBM-S model is different from the other models in that it
encodes an initial superposition of two Gaussians, turning into a single
Gaussian at long times. We note that while the short-time exponential shape
may point towards a universal, extreme-value jump-dominated dynamics \cite{eli},
data also show stretched-Gaussian shapes \cite{ilpo}, as well as long(er)-time
convergence towards an exponential \cite{andrey}. Clearly, the phenomenology
of heterogeneous environments is rich and needs further investigation.

Experimentally, the behaviours unveiled here may be used to explore further the
relevance of the different possible stochastic formulations of random-diffusivity
processes. For instance, in artificially crowded media one may vary the Hurst
exponent by changing the volume fraction of crowders or the tracer sizes, or
add drugs to change the system from super- to subdiffusive \cite{christine}.
Comparison of the resulting scaling behaviours of MSD and associated effective
diffusivity may then yield decisive clues.

The results found here will also be of interest in mathematical finance. In
fact, the original DD model is equivalent to the Heston model \cite{heston}
used to describe return dynamics of financial markets. Fractional Gaussian
noise in mathematical finance is used to include an increased "roughness" to
the emerging dynamics \cite{euch}. The different models studied here could
thus enrich market models.

The CLT is a central dogma in statistical physics, ba\-sed on the
fact that the entry variables are identically distributed. For inhomogeneous
environments, ubiquitous in many complex systems, new concepts generalising the
CLT will have to be developed. While random-diffusivity models are a
start in this direction and provide relevant strategies for data analyses
\cite{thapa}, ultimately more fundamental models including the quenched nature
of the disordered environment \cite{zhenya} need to be conceived.

\begin{appendix}

\section{Dimensionless units for the FBM-DD model}
\label{sec-1}

In dimensional form the starting equations governing the evolution of the position
$x(t)$ of the diffusing particle in the fractional version of the minimal DD-model
read
\begin{eqnarray} 
\label{seq-1}
\frac{d}{dt}x(t)=\sqrt{2D(t)}\sigma_1\xi_H(t),\,\,\,
D(t)=Y^2(t),\,\,\,
\frac{d}{dt}Y(t)=-\frac{Y}{\tau_c}+\sigma_2\eta(t).
\end{eqnarray}
Here $D(t)$ is the diffusion coefficient of dimension $[D]=\mathrm{cm}^2/
\mathrm{sec}$, $\xi_H$ represents fractional Gaussian noise with the Hurst
index $H\in(0,1]$ whose dimension is $[\xi_H]=\mathrm{sec}^{H-1}$ and whose
correlation function reads \cite{mandelbrot}
\begin{eqnarray}
\label{seq-2}
\langle\xi_H(t)\xi_H(t+\tau)\rangle\equiv\langle\xi^2_H\rangle_\tau=(2\delta^{2})^{
-1}\left((\tau+\delta)^{2H}-2\tau^{2H}+|\tau-\delta|^{2H}\right).
\end{eqnarray}
Moreover, $\sigma_1$ in equation (\ref{seq-1}) is the noise amplitude of dimension $[
\sigma_1]=\mathrm{sec}^{1/2-H}$. $Y(t)$ is an auxiliary Ornstein-Uhlenbeck process
with correlation time $\tau_c$, $\eta(t)$ is a white Gaussian noise with zero mean
and unit variance. $\sigma_2$ of units
$[\sigma_2]=\mathrm{cm}/\mathrm{sec}$ is the noise amplitude associated with $\eta
(t)$. To simplify the calculations and to obtain a more elegant formulation we
introduce dimensionless variables according to $t'=t/t_0$ and $x'=x/x_0$.
Equations (\ref{seq-1}) then become
\begin{eqnarray*} 
\fl\frac{dx'}{dt'}=\sqrt{2D(t_0t')}\frac{t_0\sigma_1}{x_0}\xi_H(t_0t'),\,\,\,
D(t_0t')=Y^2(t_0t'),\,\,\,
\frac{dY}{dt'}=-\frac{Y}{\tau_c/t_0}+\sigma_2t_0\eta(t_0t').
\end{eqnarray*}
Noting that for the Gaussian noise sources we have $\xi_H(t_0t')=t_0^{H-1}\xi_H(
t')$ and $\eta(t_0t')=t_0^{-1/2}\eta(t')$ we rewrite equations (\ref{seq-1}) as 
\begin{eqnarray*}
\frac{dx'}{dt'}=\sqrt{2\overline{D}(t')}\xi_H(t'),\,\,\,
\overline{D}(t')=\overline{Y}^2(t'),\,\,\,
\frac{d\overline{Y}}{dt'}=-\frac{\overline{Y}}{\overline{\tau}_c}+\overline{
\sigma}_2\eta(t'),
\end{eqnarray*}
where 
\begin{eqnarray*}
\overline{D}=\frac{\sigma_1^2t_0^{2H}}{x_0^2}D,\quad\overline{Y}=\frac{\sigma_1
t_0^H}{x_0}Y,\quad\overline{\tau}_c=\frac{\tau_c}{t_0},\quad\overline{\sigma_2}=
\frac{\sigma_1t_0^{1/2+H}}{x_0}\sigma_2.
\end{eqnarray*}
Now, we choose the temporal and spatial scales such that $\overline{\tau}_c=
\overline{\sigma}_2=1$, such that
\begin{equation}
\nonumber
t_0=\tau_c,\quad x_0=\sigma_1\sigma_2\tau_c^{1/2+H}.
\end{equation}
With this choice of units, the stochastic equations of our minimal FBM-DD
model are then given by equation (1) and (2) of the main text.

\section{Calculation of the integral kernel $K(\tau)$}
\label{sec-2}

Introducing $a(\tau)=e^{-\tau}$ and $b(\tau)=\sqrt{1-a^2(\tau)}$ we write $K(\tau)$
in equation (4) of the main text as 
\begin{eqnarray}
\label{seq-3}
\nonumber
K(\tau)&=&\langle|Y(t_1)||Y(t_2)|\rangle_{\tau=|t_2-t_1|}\\
\nonumber
&=&\int_{-\infty}^{\infty}
dY_1\int_{-\infty}^{\infty}dY_2|Y_1||Y_2|\frac{\exp\left(-[Y_1^2-2aY_1Y_2+Y_2^2]/
b^2\right)}{\pi b}\\
\nonumber
&=&\frac{2}{\pi b}\int_0^{\infty}dY_1\int_0^{\infty}dY_2Y_1Y_2\exp\left(-\frac{
Y_1^2-2aY_1Y_2+Y_2^2}{b^2}\right)\\
\nonumber
&&+\frac{2}{\pi b}\int_0^{\infty}dY_1\int_0^{\infty}
dY_2Y_1Y_2\exp\left(-\frac{Y_1^2+2aY_1Y_2+Y_2^2}{b^2}\right)\\
\nonumber
&=&\frac{b}{\pi}\int_0^{\infty}dY_1\frac{\partial}{\partial a}\left[\int_0^{\infty}
dY_2\exp\left(-\frac{Y_1^2+Y_2^2}{b^2}+\frac{2a}{b^2}Y_1Y_2\right)\right]\\
\nonumber
&&-\frac{b}{\pi}\int_0^{\infty}dY_1\frac{\partial}{\partial a}\left[\int_0^{\infty}
dY_2\exp\left(-\frac{Y_1^2+Y_2^2}{b^2}-\frac{2a}{b^2}Y_1Y_2\right)\right]\\
&=&B_1-B_2.
\end{eqnarray}
Using the integral
\begin{eqnarray*}
\int_0^{\infty}\exp\left(-px^2-qx\right)dx=\frac{1}{2}\sqrt{\frac{\pi}{p}}\exp
\left(\frac{q^2}{4p}\right)\mathrm{erfc}\left(\frac{q}{2\sqrt{p}}\right),
\end{eqnarray*}
where $\mathrm{erfc}(z)=1-\mathrm{erf}(z)=2\pi^{-1/2}\int_z^{+\infty}e^{-t^2}dt$
is the complementary error function, we rewrite $B_1$ and $B_2$ as 
\begin{eqnarray}
\nonumber
\fl B_1&=&\frac{b}{\pi}\int_0^{\infty}dY_1\exp\left(-\frac{Y_1^2}{b^2}\right)\frac{
\partial}{\partial a}\left[\frac{\sqrt{\pi }b}{2}\exp\left(\frac{a^2Y_1^2}{b^2}
\right)\mathrm{erfc}\left(-\frac{a}{b}Y_1\right)\right]\\
\fl&=&\frac{b^2}{2\sqrt{\pi}}\int_0^{\infty}dY_1\exp\left(-\frac{Y_1^2}{b^2}\right)
\frac{\partial}{\partial a}\left[\exp\left(\frac{a^2Y_1^2}{b^2}\right)\bigg(1+
\mathrm{erf}\left(\frac{a}{b}Y_1\right)\bigg)\right],
\label{seq-4}
\end{eqnarray}
and
\begin{eqnarray}
\nonumber
\fl B_2&=&\frac{b}{\pi}\int_0^{\infty}dY_1\exp\left(-\frac{Y_1^2}{b^2}\right)\frac{
\partial}{\partial a}\left[\frac{\sqrt{\pi }b}{2}\exp\left(\frac{a^2Y_1^2}{b^2}
\right)\mathrm{erfc}\left(\frac{a}{b}Y_1\right)\right]\\
\fl&=&\frac{b^2}{2\sqrt{\pi}}\int_0^{\infty}dY_1\exp\left(-\frac{Y_1^2}{b^2}\right)
\frac{\partial}{\partial a}\left[\exp\left(\frac{a^2Y_1^2}{b^2}\right)\bigg(1-
\mathrm{\mathrm{erf}}\left(\frac{a}{b}Y_1\right)\bigg)\right].
\label{seq-5}
\end{eqnarray}
Plugging equations (\ref{seq-4}) and (\ref{seq-5}) into (\ref{seq-3}) and after some
transformations, we get
\begin{eqnarray}
\nonumber
K(\tau)&=&\frac{b^2}{\sqrt{\pi}}\int_0^{\infty}dY\exp\left(-\frac{Y^2}{b^2}\right)
\frac{\partial}{\partial a}\left[\exp\left(\frac{a^2Y^2}{b^2}\right)\mathrm{erf}
\left(\frac{a}{b}Y\right)\right]\\
\nonumber
&=&\frac{2a}{\sqrt{\pi}}\int_0^{\infty}dYY^2e^{-Y^2}\mathrm{erf}\left(\frac{a}{b}Y
\right)+\frac{2b}{\pi}\int_0^{\infty}dYY\exp\left(-\frac{Y^2}{b^2}\right)\\
&=&\frac{1}{\pi}\left[b(\tau)+a(\tau)\arctan\left(\frac{a(\tau)}{b(\tau)}
\right)\right],
\label{seq-6}
\end{eqnarray}
which is equation (4) in the main text. This result is verified by simulation of the
Ornstein-Uhlenbeck process in figure S1. We immediately obtain the first-order and
second-order derivatives of $K(\tau)$ with respect to $\tau$,
\begin{eqnarray}
\label{seq-7}
K'(\tau)=-\frac{1}{\pi}a(\tau)\arctan\left(\frac{a(\tau)}{b(\tau)}\right),
\end{eqnarray}
and
\begin{eqnarray}
\label{seq-8}
K''(\tau)=\frac{1}{\pi}a(\tau)\left[\arctan\left(\frac{a(\tau)}{b(\tau)}\right)
+\frac{a(\tau)}{b(\tau)}\right].
\end{eqnarray}
$K(\tau)$, $K'(\tau)$ and $K''(\tau)$ are all monotonic and have the following
limits
\begin{eqnarray}
\nonumber
&&\lim\limits_{\tau\to 0}K(\tau)=\frac{1}{2},\quad\lim\limits_{\tau\to+\infty}K(
\tau)=\frac{1}{\pi},\\
\nonumber
&&\lim\limits_{\tau\to 0}K'(\tau)=-\frac{1}{2},\quad\lim\limits_{\tau\to+\infty}
K'(\tau)=0,\\
&&\lim\limits_{\tau\to 0}K''(\tau)=+\infty,\quad\lim\limits_{\tau\to+\infty}K
''(\tau)=0.
\label{seq-9}
\end{eqnarray}

\begin{figure}
\centering
\includegraphics[width=6cm]{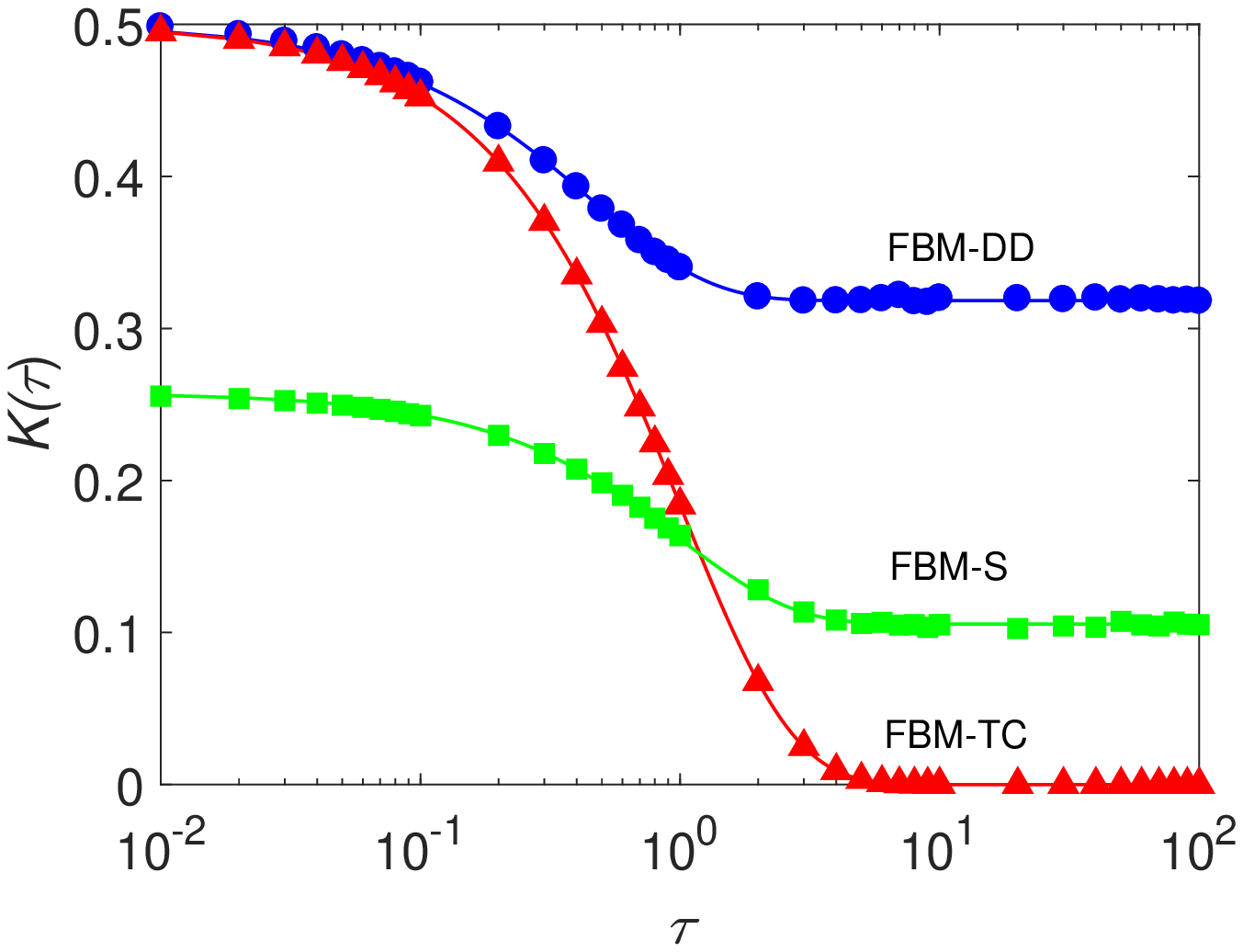}
\caption{Comparison between simulations of the respective stochastic equations
(circles, triangles, and squares) and the exact autocorrelation function $K(\tau)$
(solid curves) of the three diffusing-diffusivity models: FBM-DD (blue,
equation (\ref{seq-6})), FBM-TC (red, equation (\ref{kernel_tc})), and FBM-S (green,
equation (\ref{seq-54})).
The parameters of the FBM-S model are $D_1=1$, $D_2=0.01$, $k_{12}=3/4$, and
$k_{21}=1/4$.}
\label{figs1}
\end{figure}

\section{Exact MSD for $H=1$ in the FBM-DD model}
\label{sec-3}

Here we derive the formula for the MSD of the FBM-DD model in the
fully persistent limit $H=1$. To this end we use equation (4) of the main text
and thus $\langle\xi_H^2\rangle=1$. As result we get
\begin{eqnarray}
\label{seq-10}
\langle x^2(t)\rangle=\frac{4}{\pi}\int_0^td\tau(t-\tau)\left[b(\tau)+a(\tau)
\arctan\left(\frac{a(\tau)}{b(\tau)}\right)\right]=\frac{4}{\pi}(tI_1-I_2),
\end{eqnarray}
where
\begin{eqnarray}
\label{seq-11}
I_1=\int_0^td\tau\left[b(\tau)+a(\tau)\arctan\left(\frac{a(\tau)}{b(\tau)}\right)
\right]
\end{eqnarray}
and
\begin{eqnarray}
\label{seq-12}
I_2=\int_0^td\tau\tau\left[b(\tau)+a(\tau)\arctan\left(\frac{a(\tau)}{b(\tau)}
\right)\right].
\end{eqnarray}
We first concentrate on $I_1$. Introducing the new variable $\varphi$ such that
$a(\tau)=\exp(-\tau)=\sin\varphi$ and $b(\tau)=\cos\varphi$, we see that $\tau=0$
corresponds to $\varphi=\pi/2$, while $\tau=\infty$ corresponds to $\varphi=0$.
With $d\tau=-\cos\varphi d\varphi/\sin\varphi$ we find
\begin{equation}
\label{seq-13}
\fl I_1=\int_{\varphi_t}^{\pi/2}d\varphi\frac{\cos\varphi}{\sin\varphi}\left[\cos
\varphi+\sin\varphi\arctan\left(\tan\varphi\right)\right]=\int_{\varphi_t}^{\pi/2}
d\varphi\left[\frac{\cos^2\varphi}{\sin\varphi}+\varphi\cos\varphi\right],
\end{equation}
where $\varphi_t=\arcsin(\exp(-\tau))$ with $\varphi_t\in(0,\pi/2)$. By using
formula 1.5.6.15 from \cite{prud} we finally obtain
\begin{eqnarray}
\label{seq-14}
I_1=\frac{\pi}{2}-2\cos\varphi_t-\ln\left(\tan\frac{\varphi_t}{2}\right)
-\varphi_t\sin\varphi_t.
\end{eqnarray}
Now we turn to the integral $I_2$ in equation (\ref{seq-12}). Introducing the indefinite
integral
\begin{eqnarray*} 
\nonumber
F(\tau)&=&\int d\tau\left[b(\tau)+a(\tau)\arctan\left(\frac{a(\tau)}{b(\tau)}
\right)\right]\\
\nonumber
&=&\int d\varphi\left[\frac{\cos^2\varphi}{\sin\varphi}+\varphi\cos\varphi\right]\\
&=&-2\cos\varphi-\ln\left|\mathrm{tan}\frac{\varphi}{2}\right|-\varphi\sin
\varphi
\end{eqnarray*} 
and integrating $I_2$ by parts yields
\begin{equation}
\label{seq-15}
\fl I_2=\tau F(\tau)\Big|_{\tau=0}^{\tau=t}-\int_0^tF(\tau)d\tau=-t\Bigg(2\cos\varphi_t
+\ln\left(\tan\frac{\varphi_t}{2}\right)+\varphi_t\sin\varphi_t\Bigg)+I_{21}
+I_{22}+I_{23},
\end{equation}
where 
\begin{eqnarray}
\label{seq-16}
I_{21}=2\int_{\varphi_t}^{\pi/2}\frac{\cos^2\varphi}{\sin\varphi}=-2\cos\varphi_t
-2\ln\left(\tan\frac{\varphi_t}{2}\right),
\end{eqnarray}
\begin{eqnarray}
\label{seq-17}
I_{22}=\int_{\varphi_t}^{\pi/2}\varphi\cos\varphi d\varphi=\frac{\pi}{2}-\cos
\varphi_t-\varphi_t\sin\varphi_t,
\end{eqnarray}
and
\begin{eqnarray}
\label{seq-18}
I_{23}=\int_{\varphi_t}^{\pi/2}d\varphi\frac{\cos\varphi}{\sin\varphi}\ln
\left(\tan\frac{\varphi}{2}\right).
\end{eqnarray}
Introducing the new variables $y=\tan\frac{\varphi}{2}$ and $z=\tan^2\frac{
\varphi}{2}$, our integral $I_{23}$ becomes
\begin{eqnarray}
\nonumber
I_{23}&=&\int_{\tan\left(\varphi_t/2\right)}^{1}\frac{1-y^2}{(1+y^2)y}\ln
ydy=\int_{\tan\left(\varphi_t/2\right)}^{1}\frac{\ln y}{y}dy-\int_{\tan
(\varphi_t/2)}^1\frac{2y\ln y}{1+y^2}dy\\
&=&-\frac{1}{2}\ln^2\left(\tan\frac{\varphi_t}{2}\right)-\frac{1}{2}\int_{
\tan^2(\varphi_t/2}^1\frac{\ln z}{1+z}dz.
\label{seq-19}
\end{eqnarray}
Taking into account formula 1.6.3.8  from \cite{prud} for the indefinite integral,
\begin{eqnarray*}
\int\frac{\ln x}{x+a}dx=\ln x\ln\frac{x+a}{a}+\mathrm{Li}_2(-\frac{x}{a}),
\end{eqnarray*}
where the polylogarithm is defined as 
\begin{eqnarray*}
\mathrm{Li}_s(z)=\sum_{k=1}^{\infty}\frac{z^k}{k^s},\quad |z|<1.
\end{eqnarray*}
For the integral $I_{23}$ in equation (\ref{seq-17}) we obtain
\begin{eqnarray}
\nonumber
I_{23}&=&-\frac{1}{2}\mathrm{Li}_2(-1)-\frac{1}{2}\ln^2\left(\tan\frac{\varphi_t}{2}
\right)+\ln\left(\tan\frac{\varphi_t}{2}\right)\ln\left(1+\tan^2\frac{\varphi_t}{2}
\right)\\
&&+\frac{1}{2}\mathrm{Li}_2\left(-\tan^2\frac{\varphi_t}{2}\right).
\label{seq-20}
\end{eqnarray}
After replacing $\mathrm{Li}_2(-1)=-\pi^2/12$ and plugging equations (\ref{seq-16}),
(\ref{seq-17}), and (\ref{seq-20}) into (\ref{seq-15}) we get
\begin{eqnarray}
\nonumber
\fl I_2=&-&t\left(2\cos\varphi_t+\ln\left(\tan\frac{\varphi_t}{2}\right)+
\varphi_t\sin\varphi_t\right)-3\cos\varphi_t-2\ln\left(\tan\frac{\varphi_t}{
2}\right)+\frac{\pi}{2}-\varphi_t\sin\varphi_t\\
\fl &+&\frac{\pi^2}{24}-\frac{1}{2}\ln^2\left(\tan\frac{\varphi_t}{2}\right)
+\ln\left(\tan\frac{\varphi_t}{2}\right)\ln\left(1+\tan^2\frac{
\varphi_t}{2}\right)+\frac{1}{2}\mathrm{Li}_2\left(-\tan^2\frac{\varphi_t}{2}\right).
\label{seq-21}
\end{eqnarray}
Now, with $\varphi_t=\arcsin(\exp(-t))=\arcsin a(t)$, $\varphi_t\in(0,\pi/2)$,
$\cos\varphi_t=b(t)$, and $\tan(\varphi_t/2)=\sin\varphi_t/(1+\cos\varphi_t)=a(t)
/(1+b(t))$, equation (\ref{seq-21}) along with (\ref{seq-14}) yields the MSD,
equation (\ref{seq-10}) in the form
\begin{eqnarray}
\nonumber
\fl \langle x^2(t)\rangle&=&\frac{2}{\pi}\Bigg[t^2+(\pi-4+2\ln2)t+6b(t)+2a(t)
\arctan\left(\frac{a(t)}{b(t)}\right)-\ln^2\big(1+b(t)\big)\\
\fl&+&2(\ln2-2)\ln\big(1+b(t)\big)-\mathrm{Li}_2\left(-\frac{1-b(t)}{1+b(t)}
\right)-\pi(1+\frac{\pi}{12})\Bigg].
\label{seq-22}
\end{eqnarray}

\section{Effective long time diffusivity $D_{\mathrm{eff}}$ for $H\in(0,1/2]$ in
the FBM-DD model}
\label{sec-4}

Consider the integral
\begin{eqnarray}
\label{seq-23}
W(0,t,\delta)=2\int_0^tK(\tau)\left\langle\xi_H^2\right\rangle_\tau d\tau
\end{eqnarray}
with $t\gg 1\gg \delta$. Then the effective long-time diffusivity of the main
text, equation (7), reads
\begin{eqnarray}
\label{seq-24}
D_{\mathrm{eff}}(H\le 1/2)=\lim\limits_{\delta\to 0,t\to\infty}W(0,t,\delta).
\end{eqnarray}
For $H=1/2$ the correlation function $\langle\xi_H^2\rangle_\tau$ is reduced to
a piece-wise function, and the efficient diffusivity becomes
\begin{eqnarray}
\label{seq-25}
\fl D_{\mathrm{eff}}(H=1/2)=\lim\limits_{\delta\to0}2\int_0^{\delta}K(\tau)\frac{
\delta-\tau}{\delta^2}d\tau=\lim\limits_{\delta\to0}\int_0^{\delta}(1-\tau)\frac{
\delta-\tau}{\delta^2}d\tau=\frac{1}{2},
\end{eqnarray}
where we approximate $K(\tau)$ in equation (\ref{seq-6}) by the first-order term when
$\tau\ll1$, i.e., $K(\tau)=(1-\tau)/2+o(\tau)$.

Next we consider the efficient diffusivity for $H\in(0, 1/2)$. Introducing the
short-time scale $\Delta$, which satisfies
\begin{equation}
\label{seq-26}
\delta\ll\Delta\ll1,
\end{equation} 
we split equation (\ref{seq-23}) into two parts,
\begin{equation}
\label{seq-27}
W(0,t,\delta)=W(0,\Delta,\delta)+W(\Delta,t,\delta).
\end{equation}
Noting that the integral variable satisfies $\tau\le\Delta\ll1$ in the first part,
we use the first-order approximate $K(\tau)=(1-\tau)/2+o(\tau)$, such that
\begin{equation}
\label{seq-28}
\fl W(0,\Delta,\delta)=H\Delta^{2H-1}-(H-\frac{1}{2})\Delta^{2H}-\frac{\delta^{2H}}{
(2H+1)(2H+2)}+o(\delta^{2H}).
\end{equation}
In the second part the integral variable satisfies $\tau\ge\Delta\gg\delta$, and
we use $\langle\xi_H^2\rangle_\tau=H(2H-1)\tau^{2H-2}$, yielding
\begin{eqnarray}
\nonumber
W(\Delta,t,\delta)&=&2HK(\tau)\tau^{2H-1}\Big|^t_\Delta-K'(\tau)\tau^{2H}\Big|^t_
\Delta+\int_{\Delta}^t\tau^{2H}K'(\tau)d\tau\\
\nonumber
&=&\frac{2H}{\pi}t^{2H-1}-H(1-\Delta)\Delta^{2H-1}+K'(\Delta)\Delta^{2H}\\
&&+\int_{\Delta}^t\tau^{2H}K''(\tau)d\tau+o(\Delta^{2H}),
\label{seq-29}
\end{eqnarray}
where $K(\Delta)=(1-\Delta)/2+o(\Delta)$ and $K'(t)\sim\exp(-t)$ for $t\gg1$.
After plugging equations (\ref{seq-28}) and (\ref{seq-29}) into (\ref{seq-27}),
we have 
\begin{eqnarray}
\nonumber
W(0,t,\delta)&=&(\frac{1}{2}+K'(\Delta))\Delta^{2H}+\int_{\Delta}^t\tau^{2H}K''
(\tau)d\tau-\frac{\delta^{2H}}{(2H+1)(2H+2)}\\
&&+\frac{2H}{\pi}t^{2H-1}+o(\Delta^{
2H})+o(\delta^{2H}).
\label{seq-30}
\end{eqnarray}
From the properties of $K(\tau)$ in equation (\ref{seq-9}), $\lim\limits_{\tau\to 0}K''
(\tau)\sim\tau^{-1/2}$ and thus $\lim\limits_{\Delta\to0}\int_{\Delta}^{t}\tau^{2H}
K''(\tau)d\tau$ converges. We then have 
\begin{equation}
\label{seq-31}
W(0,t,\delta)\sim\int_0^t\tau^{2H}K''(\tau)d\tau-\frac{\delta^{2H}}{(2H+1)(2H+2)}
+\frac{2H}{\pi}t^{2H-1}.
\end{equation}
Considering the definition of the effective diffusivity, equation (\ref{seq-24}), and
combining with the case $H=1/2$ we get
\begin{equation}
\label{seq-32}
D_{\mathrm{eff}}(H\le 1/2)=\left\{\begin{array}{ll}
\int_0^{\infty}\tau^{2H}K''(\tau)d\tau,&0<H<1/2\\[0.2cm]
1/2,&H=1/2\\\end{array}\right..
\end{equation}
The long-time effective diffusivity approaches $1/2$ when $H\to0$ as $\lim\limits_{
H\to0}\int_0^{\infty}K''(\tau)d\tau=1/2$ and is discontinuous at $H=1/2$ because
$\lim\limits_{H\to1/2}\int_0^{\infty}\tau^{2H}K''(\tau)d\tau=1/2-1/\pi$.

\section{FBM-generalisation of the Tyagi-Cherayil model}
\label{sec-5}

We now consider the fractional Tyagi-Cherayil (TC) model
\label{seq-33}
\begin{eqnarray} 
\frac{dx}{dt}&=&\sqrt{2}Z(t)\sigma_1\xi_H(t),\\
\frac{dZ}{dt}&=&-\frac{Z(t)}{\tau_c}+\sigma_2\eta(t).  
\end{eqnarray}
Here $\xi_H(t)$ represents fractional Gaussian noise, $\eta(t)$ is a white Gaussian
noise, and the respective correlation functions are the same as in equation
(\ref{seq-1}).
$Z(t)$ has dimension $[Z(t)]=\mathrm{cm}/\mathrm{sec}^{1/2}$ and $[\sigma_1]=\mathrm{
sec}^{1/2-H}$, $[\sigma_2]=\mathrm{cm}/\mathrm{sec}$.

Equation (\ref{seq-34}) can be solved analytically,
\begin{eqnarray}
\label{seq-35}
\langle x^2(t)\rangle=tM_1-M_2+M_3,
\end{eqnarray}
where 
\label{seq-36}
\begin{eqnarray}
\nonumber
M_1&=&\frac{1}{\delta^2}\left(e^{\delta}\gamma(2H+1,t+\delta)-2\gamma(2H+1,t)+e^{
-\delta}\gamma(2H+1,t-\delta)\right.\\
&&-\left.e^\delta\gamma(2H+1,\delta)+e^{-\delta}\int_0^{\delta}e^xx^{2H}dx\right),
\end{eqnarray}
\begin{eqnarray}
\nonumber
M_2&=&\frac{1}{\delta^2}\Big(e^{\delta}\gamma(2H+2,t+\delta)-2\gamma(2H+2,t)+e^{
-\delta}\gamma(2H+2,t-\delta)\\
&&-e^\delta\gamma(2H+2,\delta)-e^{-\delta}\int_0^{\delta}e^xx^{2H+1}dx\Big),
\end{eqnarray}
and
\begin{eqnarray}
\nonumber
M_3&=&\frac{1}{\delta}\Big(e^{\delta}\gamma(2H+1,t+\delta)-e^{-\delta}\gamma(2H+1,t-
\delta)-e^\delta\gamma(2H+1,\delta)\\
&&-e^{-\delta}\int_0^{\delta}e^xx^{2H}dx\Big).
\end{eqnarray}
Considering the leading term of the Taylor expansion in terms of $\delta$ we get
\label{seq-37}
\begin{equation} 
M_1=2H(e^{-t}t^{2H-1}+\gamma(2H,t))-\frac{\delta^{2H}}{(2H+1)(H+1)}+o(\delta^{2H}),
\end{equation}
\begin{equation}  
M_2=(2H+1)\left(e^{-t}t^{2H}+\gamma(2H+1,t)\right)-\frac{\delta^{2H}}{H+1}+o(
\delta^{2H}),
\end{equation}
and
\begin{equation}  
M_3=2\left(e^{-t}t^{2H}+\gamma(2H+1,t)\right)-\frac{2\delta^{2H}}{2H+1}+o(
\delta^{2H}).
\end{equation}
After plugging equation (\ref{seq-37}) into (\ref{seq-35}) we get
\begin{equation}
\label{seq-38}
\fl\langle x^2(t)\rangle=e^{-t}t^{2H}+2H\gamma(2H,t)t+(1-2H)\gamma(2H+1,t)-\frac{
(t+1)\delta^{2H}}{(H+1)(2H+1)}+o(\delta^{2H}).
\end{equation}
At short times $t$ with $\delta\ll t\ll1$, $\gamma(a,t)=\int_0^te^{-x}x^{a-1}dx
\sim\int_0^t(1-x)x^{a-1}dx\sim t^a$, such that we have
\begin{equation}
\label{seq-39}
\langle x^2(t)\rangle\simeq t^{2H}
\end{equation}
At long times $t$ satisfying $\delta\ll1\ll t$ we have
\begin{equation}
\label{seq-40}
\langle x^2(t)\rangle\sim 2D_{\mathrm{eff}}t.
\end{equation}
Here $D_{\mathrm{eff}}$ can be calculated as
\begin{equation}
\label{seq-41}
D_{\mathrm{eff}}=\lim\limits_{\delta\to 0,t\to\infty}\frac{\langle x^2(t)\rangle}{
2t}=\frac{\Gamma(2H+1)}{2}
\end{equation}
For both persistence and anti-persistence cases, a crossover from anomalous diffusion
to normal diffusion emerges. The simple discussion of the FBM-DD model in the main
text can be applied to the FBM-TC model and we come to the same results (\ref{seq-38})
and (\ref{seq-39}). The definition of the long-time effective diffusivity (15) of the
main text coincides with equation (\ref{seq-41}). For finite, small values of $\delta$ and
large values of $t$, 
\begin{eqnarray}
\nonumber
\fl\frac{\langle x^2(t)\rangle}{2t}&=&\frac{e^{-t}t^{2H-1}}{2}+H\gamma(2H+1)+\frac{
(1-2H)\gamma(2H+1,t)}{2t}-\frac{\delta^{2H}}{(2H+1)(2H+2)}\\
\fl&\sim&\frac{\Gamma(2H+1)}{2}-\frac{\delta^{2H}}{(2H+1)(2H+2)}+\frac{(1-2H)\Gamma(
2H+1)}{2t}.
\label{seq-42}
\end{eqnarray}
The second term on the right hand side contributes to the discrepancies near $H\to0$
in figure 2(b) of the main text.

We expect the same behaviour of the PDF as for the DD model of Ref.~\cite{prx} but
with the rules of FBM. In particular, at short times we
expect the superstatistical behaviour to hold and the PDF should be given by the
weighted average of a single Gaussian over the stationary diffusivity distribution
of the OU process. Therefore the expected PDF reads
\begin{eqnarray}
\nonumber
\fl P(x,t)&=&\int_{-\infty}^{+\infty}p_Z(Z)G(2Z^2t^{2H})dZ\\
\nonumber
&=&\int_{-\infty}^{\infty}
\frac{1}{\sqrt{\pi}}\exp\left(-Z^2\right)\frac{1}{\sqrt{4\pi Z^2t^{2H}}}\exp\left(
-\frac{x^2}{4Z^2t^{2H}}\right)dZ\\
\fl&=&\frac{1}{2\pi t^H}\int_0^{\infty}\frac{1}{s}\exp\left(-s-\frac{x^2}{4st^{2H}}
\right)ds=\frac{1}{\pi t^H}K_0\left(\frac{x}{t^H}\right),
\label{seq-43}
\end{eqnarray}
where $G(\sigma^2)=(2\pi\sigma^2)^{-1/2}\exp(-x^2/(2\sigma^2))$ is the Gaussian
distribution, $p_Z(Z)$ is the PDF of the dimensionless OU-process, and $K_0$ is
the modified Bessel function of the second kind. At longer times the Gaussian
limit will be reached,
\begin{equation}
\label{seq-44}
P(x,t)=G(\Gamma(2H+1)t).
\end{equation}
In particular, for $H=1$, the PDF is always exponential at both short and long
times.

This can be seen from examination of the kurtosis, namely, the fourth
order moment of the displacement reads
\begin{eqnarray}
\nonumber
\langle x^4(t)\rangle&=&\int_0^tds_1\int_0^tds_2\int_0^tds_3\int_0^tds_4\langle
D(s_1)D(s_2)D(s_3)D(s_4)\rangle\\
\nonumber
&&\times \langle\xi_H(s_1)\xi_H(s_2)\xi_H(s_3)\xi_H(s_4)
\rangle\\
\nonumber
&=&3\int_0^tds_1\int_0^tds_2\int_0^tds_3\int_0^tds_4\langle D(s_1)D(s_2)D(s_3)
D(s_4)\rangle\\
&&\times\langle\xi_H(s_1)\xi_H(s_2)\rangle\langle\xi_H(s_3)\xi_H(s_4)\rangle.
\label{seq-45}
\end{eqnarray}
For $H=1$, $\langle\xi^2_H\rangle_\tau=1$ and the forth moment becomes
\begin{eqnarray}
\nonumber
\langle x^4(t)\rangle&=&3\int_0^t ds_1\int_0^tds_2\int_0^tds_3\int_0^tds_4\langle
D(s_1)D(s_2)D(s_3)D(s_4)\rangle\\
\nonumber
&=&9\int_0^tds_1\int_0^tds_2\langle D(s_1)D(s_2)\rangle\int_0^tds_3\int_0^tds_4
\langle D(s_3)D(s_4)\rangle\\
&=&9\langle x^2(t)\rangle^2
\label{seq-46}
\end{eqnarray}
Thus the kurtosis for $H=1$ reads
\begin{equation}
\label{seq-47}
k=\frac{\langle x^4(t)\rangle}{\langle x^2(t)\rangle^2}=9
\end{equation}
This means that for $H=1$, the crossover to the Gaussian will never emerge at any
time. This is a fundamental distinction from the FBM-DD model. The behaviour
of the kurtosis is shown in figure S2.

\section{FBM-Switching model}
\label{sec-6}

Due to the Markovian nature of the S-model (\ref{swit}), the  matrix of the transition
probabilities of $n(t)$ is ($\mathrm{Pr}$ denotes probability)
\begin{equation}
\label{seq-49}
\fl
\mathrm{Pr}\{n(t)=i|n(0)=j\}=\tau_c\left(\begin{array}{cc}k_{21}+k_{12}e^{-t/\tau_c}
& k_{21}(1-e^{-t/\tau_c})\\ k_{12}(1-e^{-t/\tau_c}) & k_{12}+k_{21}e^{-t/\tau_c}
\end{array}\right),\quad i,j=0,1.
\end{equation}
The stationary probability of $n(t)$ is 
\begin{equation}
\label{seq-50}
\mathrm{Pr}\{n(t)=0\}=k_{21}\tau_c,\quad\mathrm{Pr}\{n(t)=1\}=k_{12}\tau_c.
\end{equation}
The mean of $n(t)$ with stationary initial condition will be
\begin{equation}
\label{seq-51}
\langle n(t)\rangle=k_{12}\tau_c,
\end{equation}
and the correlation function becomes
\begin{eqnarray}
\nonumber
\fl\langle n(t)n(t')\rangle&=&\mathrm{Pr}\{n(t)=1,n(t')=1\}=\mathrm{Pr}\{n(t')=1
|n(t)=1\}\times\mathrm{Pr}\{n(t)=1\}\\
\fl&=&(k_{12}+k_{21}e^{-\tau/\tau_c})k_{12}\tau_c^2.
\label{seq-52}
\end{eqnarray}
Using equation (\ref{swit}) we obtain the first and second moments of $\theta$,
\begin{equation}
\label{seq-53}
\langle\theta(t)\rangle=(D_2^{1/2}-D_1^{1/2})\langle n(t)\rangle+D_1^{1/2}=
(k_{21}D_1^{1/2}+k_{12}D_2^{1/2})\tau_c
\end{equation}
and
\begin{equation}
\label{seq-54}
\fl\langle\theta(t)\theta(t')\rangle=(D_2^{1/2}-D_1^{1/2})^2\langle n(t)n(t')
\rangle+2D_1^{1/2}(D_2^{1/2}-D_1^{1/2})\langle n(t)\rangle+D_1=a_1e^{-\tau/\tau_c}
+a_2.
\end{equation}
Here, $\tau=|t-t'|$, $a_1=(D_2^{1/2}-D_1^{1/2})^2k_{12}k_{21}\tau_c^2$, and $a_2
=(k_{21}D_1^{1/2}+k_{12}D_2^{1/2})^2\tau_c^2$. The correlation (shown in figure S1
in comparison to Langevin simulations) approaches $a_2+a_2$ at short times and
$a_2$ at long times. 

For finite values $\delta$ and $t$ in the persistent case ($H>1/2$), we find the MDS
\begin{eqnarray}
\nonumber
\fl\frac{\langle x^2(t)\rangle}{2t^{2H}}&=&a_1e^{-t/\tau_c}+a_2+\left[2Ha_1\tau_c
^{2H-1}\gamma(2H,t/\tau_c)-\frac{a_1\delta^{2H}}{(H+1)(2H+1)}\right]t^{1-2H}\\
\nonumber
\fl&&+(1-2H)a_1\tau_c^{2H}\gamma(2H+1,t/\tau_c)t^{-2H}\\
\fl&\sim&a_2+\Gamma(2H+1)a_1\tau_c^{2H-1}t^{1-2H},
\label{seq-61}
\end{eqnarray}
while in the anti-persistent case ($H<1/2$),
\begin{eqnarray}
\nonumber
\frac{\langle x^2(t)\rangle}{2t}&=&a_1e^{-t/\tau_c}t^{2H-1}+a_2t^{2H-1}+2Ha_1
\tau_c^{2H-1}\gamma(2H,t/\tau_c)\\
\nonumber
&&+(1-2H)a_1\tau_c^{2H}\gamma(2H+1,t/\tau_c)t^{-1}
-\frac{a_1\delta^{2H}}{(H+1)(2H+1)}\\
&\sim&\Gamma(2H+1)a_1\tau^{2H-1}_c+a_2t^{2H-1}-\frac{a_1\delta^{2H}}{(H+1)(2H+1)}.
\label{seq-62}
\end{eqnarray}

The fourth oder moment of the displacement reads
\begin{eqnarray}
\nonumber
\langle x^4(t)\rangle&=&4\int_0^tds_1\int_0^tds_2\int_0^tds_3\int_0^tds_4\langle
\theta(s_1)\theta(s_2)\theta(s_3)\theta(s_4)\rangle\\
\nonumber
&&\times\langle\xi_H(s_1)\xi_H(s_2)\xi
_H(s_3)\xi_H(s_4)\rangle\\
\nonumber
&=&12\int_0^tds_1\int_0^tds_2\int_0^tds_3\int_0^tds_4\langle\theta(s_1)\theta(s_2)
\theta(s_3)\theta(s_4)\rangle\\
&&\times\langle \xi_H(s_1)\xi_H(s_2)\rangle\langle\xi_H(s_3)
\xi_H(s_4)\rangle.
\label{seq-65}
\end{eqnarray}
At short times, $\langle\theta(s_1)\theta(s_2)\theta(s_3)\theta(s_4)\rangle\approx
\langle\theta^4(t)\rangle=\langle\theta^4(t)\rangle=(k_{21}D_1^2+k_{12}D_2^2)\tau_c$.
With equation (\ref{seq-65}) the kurtosis reads
\begin{eqnarray}
\label{seq-66}
k=\frac{\langle x^4(t)\rangle}{\langle x^2(t)\rangle^2}=\frac{3(k_{21}D_1^2+k_{12}
D_2^2)}{(k_{21}D_1+k_{12}D_2)^2\tau_c}.
\end{eqnarray}

The behaviours of the kurtosis of the three different random-diffusivity models
are shown in figure S2.

\begin{figure*}
\includegraphics[width=5.2cm]{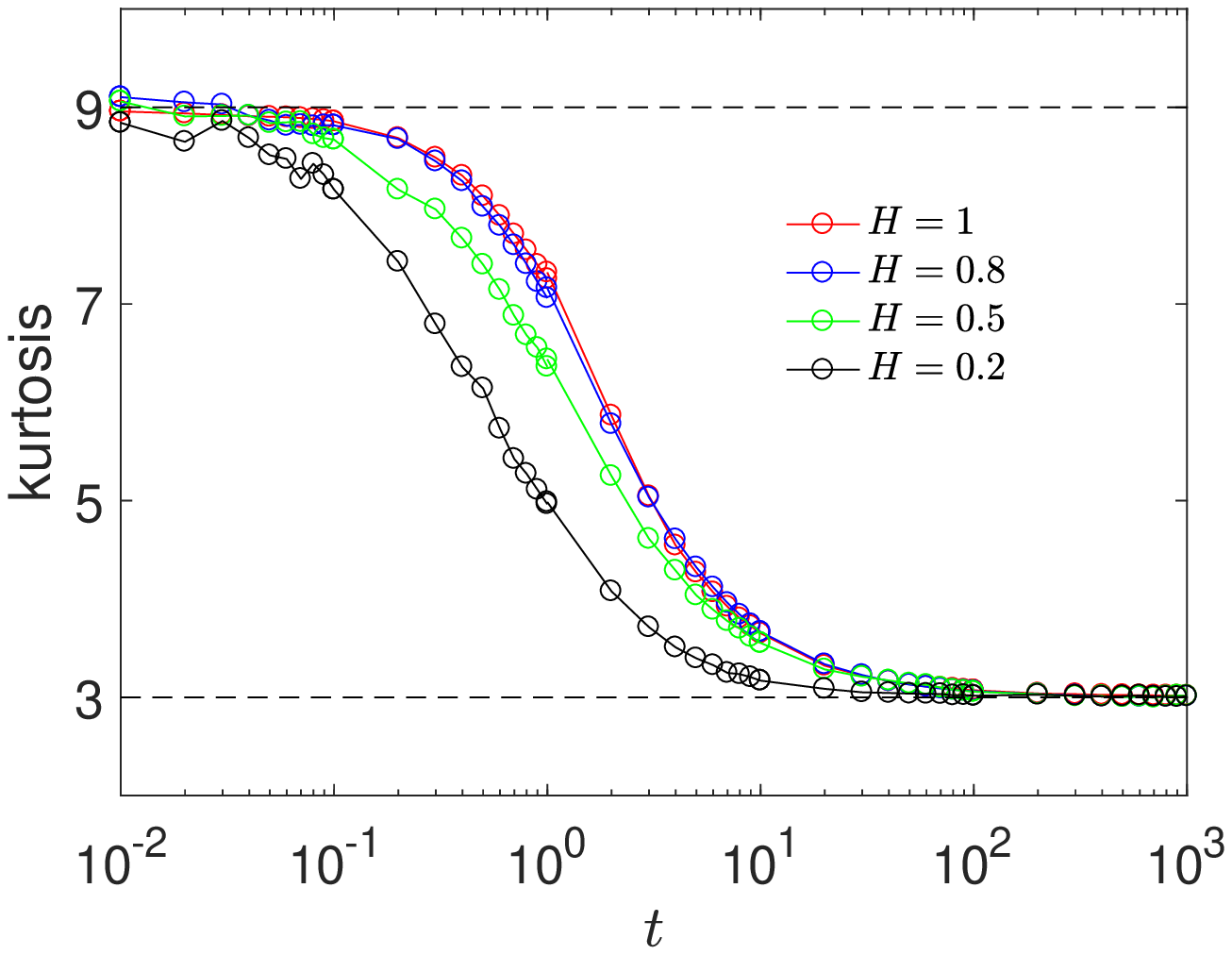}
\includegraphics[width=5.2cm]{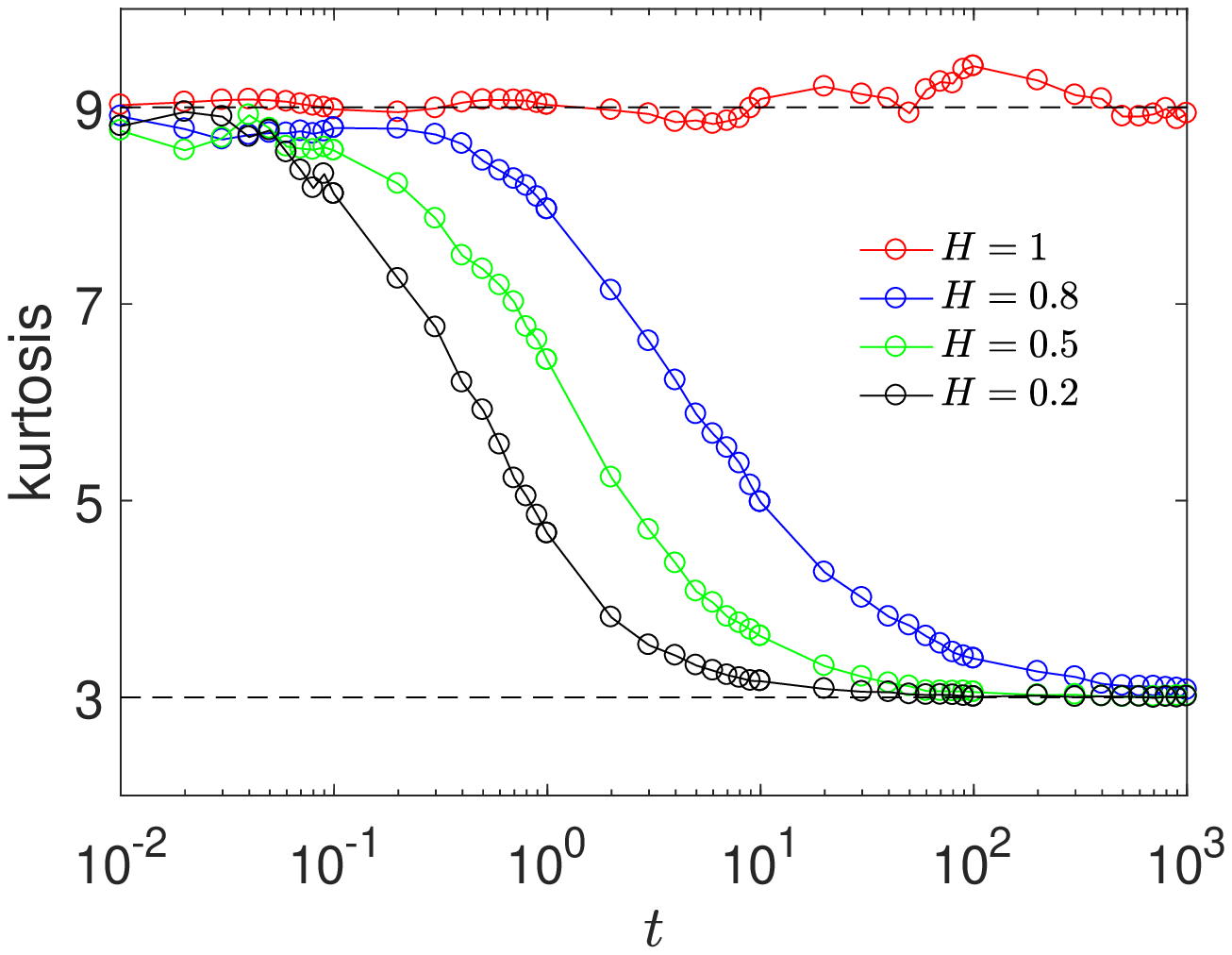}
\includegraphics[width=5.2cm]{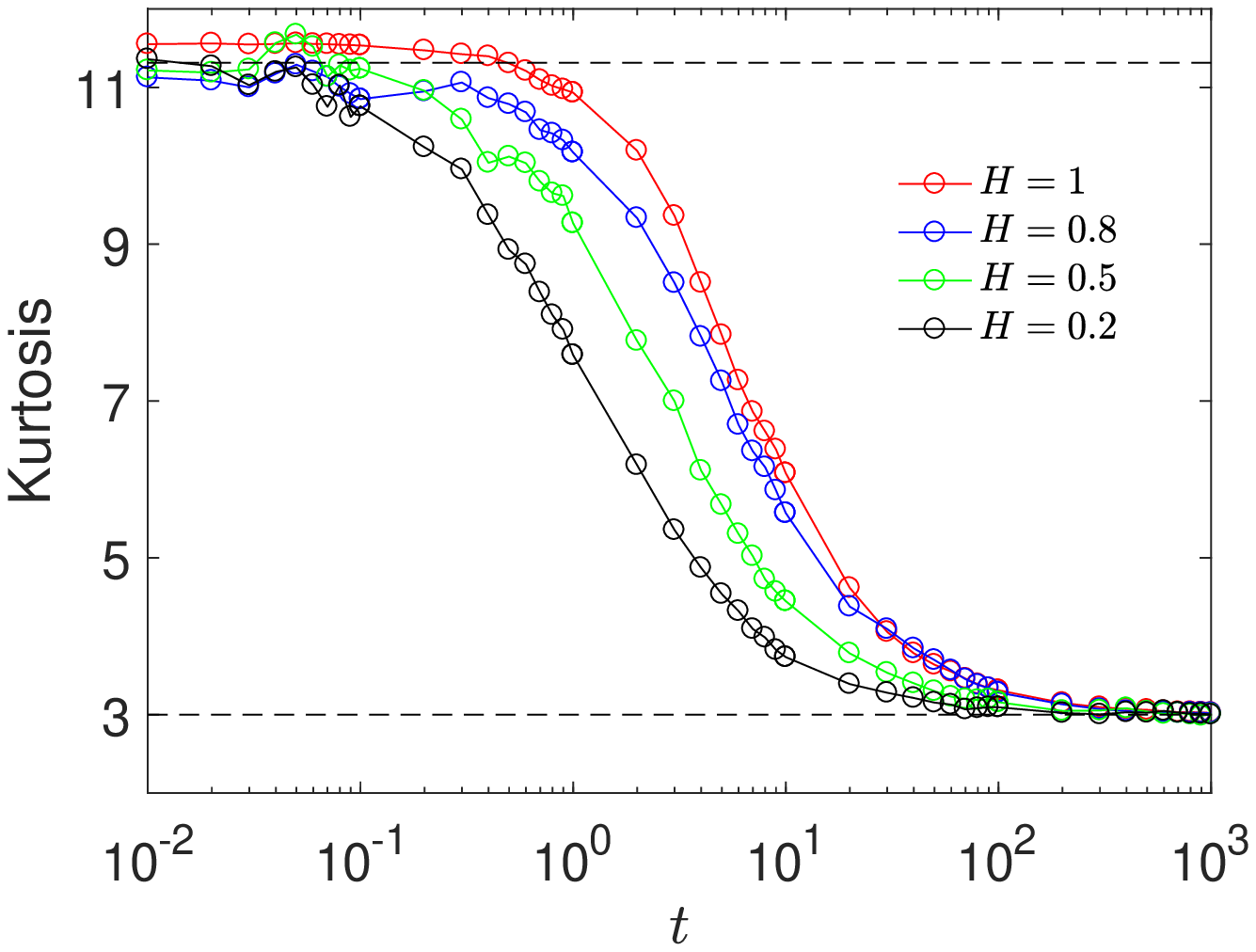}
\caption{Langevin simulations of the kurtosis for the three random-diffusivity
models: (a) FBM-DD, (b) FBM-TC, (c) FBM-S. Parameters of the FBM-S model: $D_1=1$,
$D_2=0.01$, $k_{12}=3/4$, and $k_{21}=1/4$.}
\end{figure*}

\end{appendix}

\ack

We acknowledge funding from DFG (ME 1535/7-1). RM acknowledges the Foundation
for Polish Science (Fundacja na rzecz Nauki Polskiej, FNP) for an Alexander
von Humboldt Polish Honorary Research Scholarship. FS acknowledges Davide
Straziota for helpful discussions and financial support of the 191017 BIRD-PRD
project of the Department of Physics and Astronomy of Padua University.

\end{document}